\documentclass[prb,twocolumn,superscriptaddress]{revtex4-1}

\pdfoutput=1

\usepackage{graphicx}
\usepackage{amsmath,amssymb}

\usepackage{color}
\usepackage{amssymb}
\usepackage{braket}

\usepackage{hyperref}
\hypersetup{colorlinks=true,breaklinks,linkcolor=blue,urlcolor=blue,citecolor=blue}

\def\mathi{\mathrm i}

\newcommand{\bS}{\ensuremath{\boldsymbol{S}}}
\newcommand{\bU}{\ensuremath{\boldsymbol{U}}}
\newcommand{\bV}{\ensuremath{\boldsymbol{V}}}

\newcommand{\bK}{\ensuremath{\boldsymbol{K}}}

\newcommand{\wmax}{\ensuremath{{\omega_\mathrm{max}}}}

\newcommand{\Gthree}{\ensuremath{G^\mathrm{3pt}}}
\newcommand{\Gthreenorm}{\ensuremath{G_\mathrm{normal}^\mathrm{3pt}}}
\newcommand{\Gthreesingular}{\ensuremath{G_\mathrm{singular}^\mathrm{3pt}}}
\newcommand{\Gfour}{\ensuremath{G^\mathrm{4pt}}}

\newcommand{\KF}{\ensuremath{K^\mathrm{F}}}
\newcommand{\KB}{\ensuremath{K^\mathrm{B}}}
\newcommand{\UF}{\ensuremath{U^\mathrm{F}}}
\newcommand{\UB}{\ensuremath{U^\mathrm{B}}}
\newcommand{\enhUB}{\ensuremath{U^\mathrm{\overline{B}}}}

\newcommand{\overlineB}{\ensuremath{{\overline{\mathrm{B}}}}}

\begin{document}
\title{Overcomplete compact representation of two-particle Green's functions}
\author{Hiroshi Shinaoka}
\affiliation{Department of Physics, Saitama University, 338-8570, Japan}

\author{Junya Otsuki}
\affiliation{Department of Physics, Tohoku University, Sendai 980-8578, Japan}

\author{Kristjan Haule}
\affiliation{Department of Physics and Astronomy, Rutgers University, Piscataway, New Jersey 08854, USA}

\author{Markus Wallerberger}
\affiliation{University of Michigan, Ann Arbor, Michigan 48109, USA}

\author{Emanuel Gull}
\affiliation{University of Michigan, Ann Arbor, Michigan 48109, USA}

\author{Kazuyoshi Yoshimi}
\affiliation{Institute for Solid State Physics, University of Tokyo, Chiba 277-8581, Japan}

\author{Masayuki Ohzeki}
\affiliation{Graduate School of Information Sciences, Tohoku University, Sendai 980-8579, Japan}

\date{\today}

\begin{abstract}
Two-particle Green's functions and the vertex functions play a critical role in theoretical frameworks for describing strongly correlated electron systems.
However, numerical calculations at two-particle level often suffer from large computation time and massive memory consumption.
We derive a general expansion formula for the two-particle Green's functions in terms of an overcomplete representation based on the recently proposed ``intermediate representation" basis.
The expansion formula is obtained by decomposing the spectral representation of the two-particle Green's function.
We demonstrate that the expansion coefficients decay exponentially, while all high-frequency and long-tail structures in the Matsubara-frequency domain are retained.
This representation therefore enables efficient treatment of two-particle quantities 
and opens a route to the application of modern many-body theories to realistic strongly correlated electron systems.
\end{abstract}

\maketitle
\section{Introduction}
Theoretical treatment of strong electronic correlations is one of the most challenging and fascinating topics in condensed matter physics.
Intensive progress has been achieved in the recent decade on vertex function based diagrammatic expansions around the dynamical mean-field theory (DMFT)~\cite{Georges96} (for a review, see Ref.~\onlinecite{Rohringer-arXiv}).
Examples include the dynamical vertex approximation~\cite{Toschi07,Schaefer15,Galler17}, the dual-fermion approach~\cite{Rubtsov09,Hafermann09,Antipov14,Otsuki14a}, and several related methods~\cite{Kusunose06,Rohringer13,Taranto14,Kitatani15,Ayral15,Ayral16,Li:2015cwa}.
All of these methods, while based on different technical formulations, aim to incorporate spatial fluctuations on top of locally correlated states constructed by the DMFT.

In correlated perturbation theories, two-particle Green's functions and vertex functions play principal roles.
Physically, they represent renormalized interactions between emerging local degrees of freedom.
Solving the Bethe-Salpeter equations yields the momentum-dependent susceptibility in DMFT~\cite{Jarrell92,Georges96,Park:2011fga,Kunes17} and improvement of the DMFT self-energy in the theories above.
Further elaborate treatments using the so-called parquet equations take into account cross-channel fluctuations between magnetism and superconductivity.\cite{Schaefer15,Gunnarsson16,Li16}

However, numerical calculations involving vertex functions are expensive.
Technical difficulties arise from complicated dependencies on three frequencies and the presence of slowly decaying high-frequency tails. 
These can contain multiple energy scales in correlated electron systems.
A recipe for describing the frequency dependence is that the vertex is computed in a low-frequency cube with a fixed frequency cutoff.
This works well at high temperature. However, as temperature is lowered, one needs a gradually larger cube, so that the low-temperature parameter regions of interest become inaccessible.
The influence of the high frequency truncation can be alleviated by considering high-frequency asymptotics of the vertex part, as examined in Refs.~\onlinecite{Kunes11,Rohringer:2012cc,Li16,Kaufmann:2017hw,Wentzell-arXiv}.

Basis transformations offer an alternative strategy. As shown in the context of single-particle Green's functions~\cite{Boehnke11}, 
orthogonal polynomial representations yield an efficient way to store single-particle Green's function. A further compact representation was recently found as an intermediate representation (IR) between Matsubara-frequency and real-frequency domains~\cite{Shinaoka17}.
Although the IR basis achieves a remarkable performance in expressing the single-particle Green's functions, its naive usage for two-particle object is problematic, as there exists non-trivial high-frequency structure in many different combinations of three frequencies. These cannot be represented by a product of orthogonal basis sets.
How to best describe the complex structure of Green's functions beyond the single-particle level is therefore an important open question.

In this paper, we present expansion formulas for two-particle Green's functions, generalizing the concept of IR to multiple-time correlation functions.
The key idea is to introduce overcomplete basis sets, which are naturally derived from the spectral representations of the two-particle Green's functions.
These expansions then capture the full high-frequency structure, so that the resulting expansion coefficients decay exponentially fast.
This representation enables the practical calculation of recent many-body theories at the two-particle level in practice.

The remainder of this paper is organized as follows.
To establish notations, we first review, in the next section, the IR basis for the single-particle Green's function.
The expansion formulas for two-particle Green's functions are derived in Sec.~\ref{sec:TPGF}.
Section~\ref{sec:coefficients} demonstrates the accuracy of the present method for a simple model.
In Section~\ref{sec:qmc}, we show numerical results for a more complicated model using data obtained by quantum Monte Carlo simulations.
Section~\ref{sec:summary} presents a summary and conclusions.

\section{Single-particle Green's function}
In this section, we review the derivation of the IR basis and see how a compact representation of the single-particle Green's functions is obtained.~\cite{Shinaoka17}
The IR basis introduced here will be used to construct a compact representation for the two-particle Green's functions in the next section.

\subsection{Spectral representation}
Let us start our discussion by considering the spectral (Lehmann) representation of the single-particle Green's function $G^{\alpha}(\tau)$ in the imaginary-time domain
\begin{align}
	G^{\alpha}(\tau) &= -\int_{-\wmax}^{\wmax} d\omega K^{\alpha}(\tau, \omega) \rho^{\alpha}(\omega),\label{eq:fwd}
\end{align}
where we take $\hbar = 1$ and $0 \le \tau  \le \beta$.
We assume that the spectrum $\rho^{\alpha}(\omega)$ is bounded in the interval $[-\wmax, \wmax]$.
The superscript $\alpha$ specifies statistics: $\alpha=\mathrm{F}$ for fermion and $\alpha=\mathrm{B}$ for boson.
The spectral function $\rho^{\alpha}(\omega)$ is given by
\begin{align}
	\rho^\alpha(\omega) &= -\frac{1}{\pi\omega^{\delta_{\alpha,\mathrm{B}} }} \mathrm{Im} G^\alpha(\omega + \mathi 0).\label{eq:rho}
\end{align}
Correspondingly, the kernel $K^{\alpha}(\tau,\omega)$ is defined by
\begin{align}
	K^\alpha(\tau, \omega) &\equiv \omega^{\delta_{\alpha,\mathrm{B}}} \frac{e^{-\tau\omega}}{1 \pm e^{-\beta \omega}}.\label{eq:K}
\end{align}
The extra $\omega$'s were introduced above to avoid the singularity of $K^\mathrm{B}(\tau,\omega)$ at $\omega=0$.\cite{MaxEnt}
The kernel has the same (anti-) periodicity as in $G(\tau)$, 
and exhibits a discontinuity at $\tau = n \beta$ ($n=0, \pm 1, \pm 2, \cdots$).

\subsection{Definition of the IR basis}
\begin{figure}
	\centering
	\includegraphics[width=0.45\textwidth]{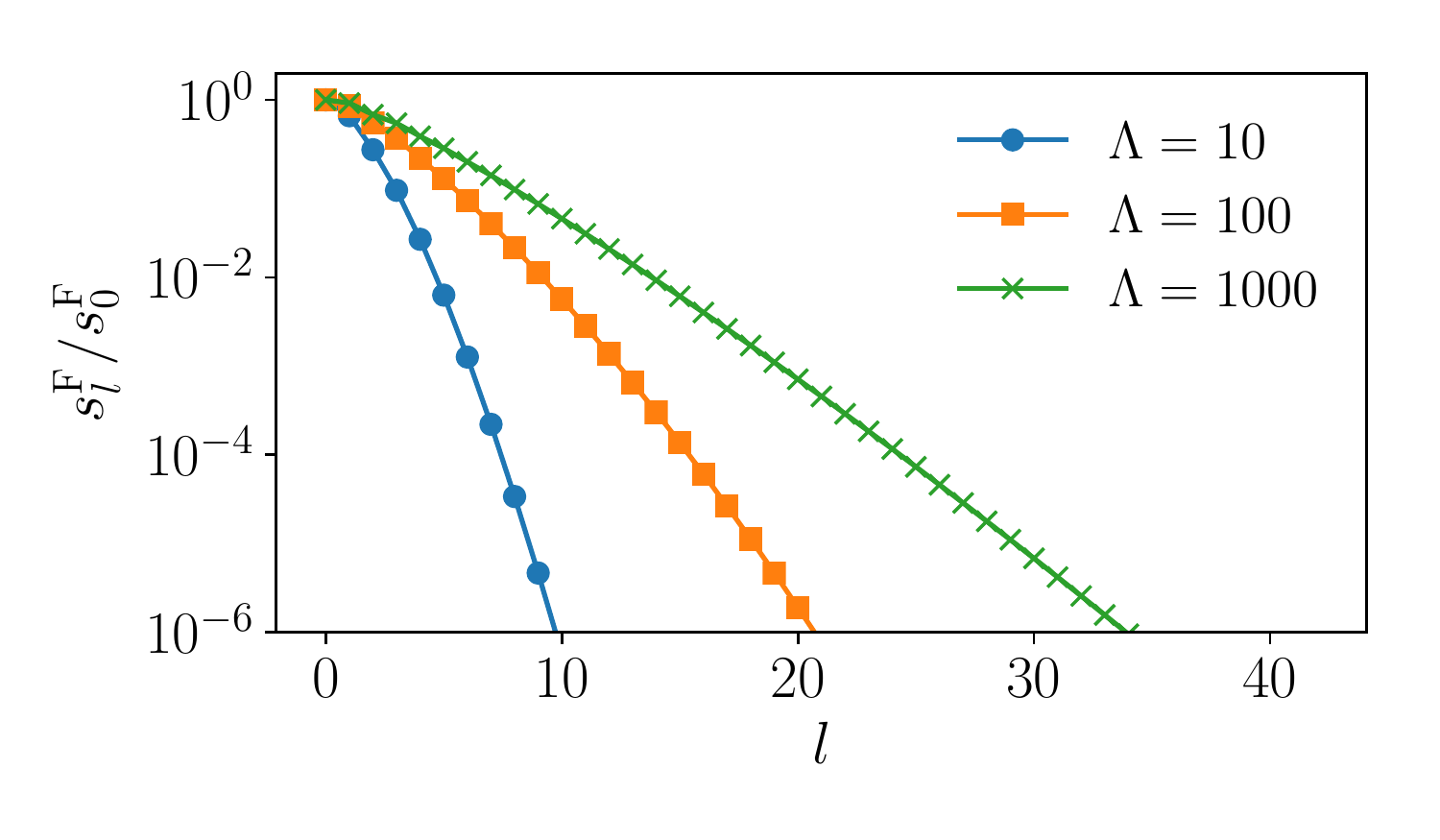}
	\caption{
		Singular values $s_l^\mathrm{F}$ for several values of $\Lambda = \beta \wmax$ [see Eq.~(\ref{eq:kernel-exp}) for the definition of $s_l^\mathrm{F}$].
	}
	\label{fig:sl}
\end{figure}

We derive two complete orthonormal basis sets, $\{U^\alpha_l(\tau)\}$ and $\{V^\alpha_l(\omega)\}$, through the decomposition
\begin{align}
K^{\alpha}(\tau, \omega) &= \sum_{l=0}^{\infty} s^{\alpha}_l U^{\alpha}_l(\tau) V^{\alpha}_l(\omega) \label{eq:kernel-exp}
\end{align}
for $\tau\in[0,\beta]$ and $\omega \in [-\wmax,\wmax]$.
These basis sets are orthogonalized as  $\int_0^\beta d \tau U_l^\alpha(\tau)U_{l^\prime}^\alpha(\tau) =\int_{-\wmax}^\wmax d \omega V_l^\alpha(\omega)V_{l^\prime}^\alpha(\omega) = \delta_{ll^\prime}$.
This decomposition corresponds to the singular value decomposition (SVD) of the kernel matrix $\bK$ defined on a discrete $\tau$-$\omega$ space: $\bK = \bU \bS \bV^T$.
Column vectors of $\bU$ and $\bV$ in the continuous limit yield $U_l(\tau)$ and $V_l(\omega)$, respectively.
We refer to these orthonormal functions as the IR basis.
We note that the IR basis and the singular values $s_l$ depend on $\beta$ and $\wmax$ through a dimensionless parameter $\Lambda=\beta \wmax$.
A striking feature of this decomposition is the exponential decay of $s_l^{\alpha}$ as depicted in Fig.~\ref{fig:sl}.

The characteristic properties of the IR basis are as follows.
For even (odd) values of $l$, $U^\alpha_l(\tau)$ and $V^\alpha_l(\omega)$ are even (odd) functions with respect to the center of the domain, i.e., $\tau=\beta/2$ or $\omega=0$.
Interestingly, $U_l^{\alpha}(\tau)$ and $V_l^{\alpha}(\omega)$ are reduced to the Legendre polynomials in the limit $\Lambda \to 0$ (if the ranges of $\tau$ and $\omega$ are scaled properly).\cite{Shinaoka17}
For $\Lambda>0$, they each constitute a non-polynomial orthogonal system.
We note that the IR basis is not well defined as $\Lambda=\infty$.
We refer readers to Ref.~\onlinecite{Chikano-unpublished} for more details on the properties of the IR basis. 

\subsection{Expansion of Green's function into the IR basis}
We expand $G^\alpha(\tau)$ using the complete basis $\{U^\alpha(\tau)\}$ as follows:
\begin{align}
G^{\alpha}(\tau) &= \sum_{l=0}^\infty  G_l^{\alpha} U_l^{\alpha}(\tau)\label{eq:IR-decomp}.
\end{align}
Substituting Eq.~(\ref{eq:kernel-exp}) into Eq.~(\ref{eq:fwd}) and comparing with the above equation, we obtain 
\begin{align}
G^{\alpha}_l \propto s^{\alpha}_l \rho^{\alpha}_l,\label{eq:gl}
\end{align}
where $\rho^{\alpha}_l$ is given by
\begin{align}
\rho_l^{\alpha} = \int_{-\wmax}^{\wmax} d \omega \rho^\alpha(\omega) V^\alpha_l(\omega).\label{eq:rhol}
\end{align}
Equation~(\ref{eq:gl}) shows that the expansion coefficients $G_l^{\alpha}$ decay at least as fast as $s_l^{\alpha}$.
Quantum Monte Carlo model calculations indeed demonstrate an exponentially fast decay of $G_l^{\alpha}$.\cite{Shinaoka17}
Furthermore, the number of basis functions for the convergence of expansion increases only logarithmically with $\beta$ (see Fig.~\ref{fig:sl}).\cite{Chikano-unpublished}

We note that the actual speed of convergence depends on the choice of $\wmax$.
In typical calculations, we set $\wmax$ to a value much larger than the spectral width of the system. This choice of cutoff value only slightly influences convergence (only logarithmically).

The Fourier transform of Eq.~(\ref{eq:IR-decomp}) yields the Matsubara Green's function as
\begin{align}
 	G^\alpha(i\omega_n) &\equiv \int_0^\beta d \tau G^\alpha(\tau) e^{i\omega_n \tau} = \sum_{l=0}^\infty　G_l^\alpha U^\alpha_l(i\omega_n)\\
	&\mathrm{with}~U^\alpha_l(i\omega_n) \equiv \int_0^\beta d \tau U^\alpha_l(\tau) e^{i\omega_n \tau}.
\end{align}

\section{Compact representation of two-particle Green's functions}
\label{sec:TPGF}

In this section, we discuss how to construct a compact representation of two-particle Green's functions.
The main difficulty in handling two-particle objects is the non-trivial high-frequency structure in many different combinations of frequencies~\cite{Kunes11,Li16,Wentzell-arXiv}.
In the imaginary-time domain, this high-frequency structure is related to imaginary time discontinuities whose origin lies in the (anti-)periodic nature of the Green's functions.
Taking those discontinuities into account properly is key to obtain exponentially decaying expansion coefficients.

In order to illustrate this challenge, we first consider a three-point Green's function in Sec.~\ref{sec:Gthree}.
The difficulty in constructing compact representations already exists there and the essence of our idea is easier to follow.
Four-point Green's functions will be addressed in Sec.~\ref{sec:Gfour}.

\subsection{The three-point Green's functions}\label{sec:Gthree}
\begin{figure}
	\centering
	\includegraphics[width=0.3\textwidth,clip]{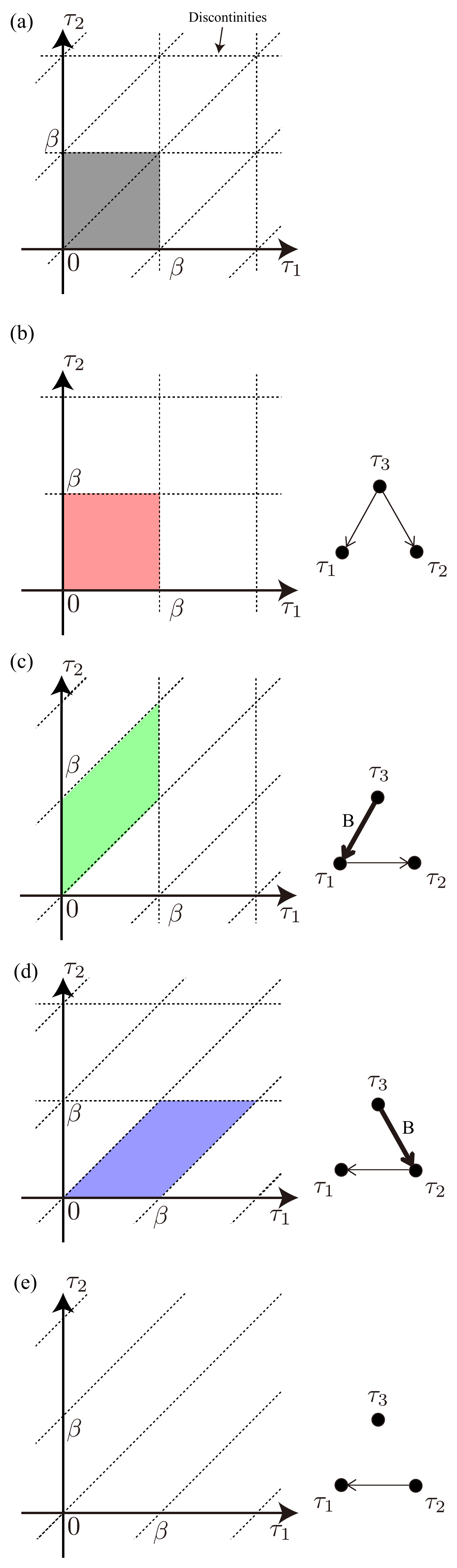}
	\caption{
		(Color online) (a) Equal-time lines of $\Gthree(\tau_1, \tau_2, \tau_3)$ in the $\tau_1$-$\tau_2$ space with $\tau_3=0$.
		(b)-(e) Decomposition of $\Gthree(\tau_1, \tau_2, \tau_3)$ into four different functions in Eq.~(\ref{eq:three-point-spectral-tau}).
		The positions of their discontinuities are represented by broken lines. The squares denoted by the shadows in (b)--(d) are a unit of (anti-) periodicity in the $\tau_1$-$\tau_2$ space.
		The corresponding representations of relative times are shown graphically.
		The thin (bold) arrow indicates fermionic (bosonic) statistics.
		The singular term shown in (e) does not depend on any relative time with bosonic statistics.
	}
	\label{fig:three-point}
\end{figure}

We consider a three-point Green's function defined by
\begin{align}
  \Gthree(\tau_1, \tau_2, \tau_3) &= \braket{T_\mathrm{\tau} A(\tau_1) B(\tau_2) C(\tau_3)},\label{eq:3point}
\end{align}
where $A$ and $B$ are fermionic operators in the Heisenberg picture and $C$ is a bosonic operator.
Figure~\ref{fig:three-point}(a) illustrates the location of discontinuities in the $\tau_1$-$\tau_2$ plane.
It turns out that there are two discontinuities in one period.

Our expansion formula for $\Gthree$ is based on its spectral representation as in the case of the single-particle Green's function.
Leaving the detailed derivation in Appendix~\ref{appendix:three-point},
we start our discussion from the final expression for the spectral represention:
\begin{align}
	 & \Gthree(\tau_1, \tau_2, \tau_3) \nonumber\\
	 & = \int_{-\infty}^\infty d\epsilon_1 d \epsilon_2 \Big\{
	 \KF(\tau_{13},\epsilon_1) \KF(\tau_{23},\epsilon_2)\rho^{(1)}(\epsilon_1,\epsilon_2)\nonumber\\
	&\quad+  \KB(\tau_{13},\epsilon_1) \KF(\tau_{21},\epsilon_2)\rho^{(2)}(\epsilon_1,\epsilon_2)\nonumber\\
	&\quad+  \KF(\tau_{12},\epsilon_1) \KB(\tau_{23},\epsilon_2)\rho^{(3)}(\epsilon_1,\epsilon_2)
	\Big\}\nonumber\\
	&\quad+ \int_{-\infty}^\infty d\epsilon
K^\mathrm{F}(\tau_{12}, \epsilon)\rho_\mathrm{singular}(\epsilon)
	,\label{eq:three-point-spectral-tau}
\end{align}
where $\tau_{ij} \equiv \tau_i - \tau_j$ denotes the relative time.
This expression consists of four terms with distinct spectral functions.
The first three correspond to three possible ways of defining pairs of relative times: $(\tau_{13}, \tau_{23})$, $(\tau_{13}, \tau_{21})$, $(\tau_{12}, \tau_{23})$.
These contributions are continuous in different domains shown in Figs.~\ref{fig:three-point}(b)--(d).
The last term in Eq.~(\ref{eq:three-point-spectral-tau}) depends only on $\tau_{12}$.
This contribution is expressed with the ``disconnected'' graph in Fig.~\ref{fig:three-point}(e).
The four terms together express all the discontinuities of $\Gthree(\tau_1, \tau_2, \tau_3)$.

Equation~(\ref{eq:three-point-spectral-tau}) indicates that $\Gthree(\tau_1, \tau_2, \tau_3)$ can be expanded using \emph{a single-particle basis} with four types of decoupling being combined.
We thus arrive at the following expansion formula for $\Gthree$:
\begin{align}
	& \Gthree(\tau_1, \tau_2, \tau_3) \nonumber\\
	& =  \sum_{l_1,l_2=0}^\infty \Big\{G^{(1)}_{l_1l_2}\UF_{l_1}(\tau_{13})\UF_{l_2}(\tau_{23}) \nonumber\\
	& +  G_{l_1 l_2}^{(2)} \UB_{l_1}(\tau_{13})\UF_{l_2}(\tau_{21}) +
	G_{l_1 l_2}^{(3)} \UF_{l_1}(\tau_{12})\UB_{l_2}(\tau_{23})
	\Big\}
	\nonumber\\
	&+ \sum_{l=0}^{\infty} g_l \UF_{l}(\tau_{12}).
\end{align}
We note that the first three terms each form a complete orthogonal basis with the different unit of (anti-) periodicities.
For a simpler description, the last term may be absorbed into the second and third terms by extending the bosonic basis as
\begin{align}
	& \Gthree(\tau_1, \tau_2, \tau_3) \nonumber\\
	& =  \sum_{l_1,l_2=0}^\infty \Big\{G^{(1)}_{l_1l_2}\UF_{l_1}(\tau_{13})\UF_{l_2}(\tau_{23}) \nonumber\\
	& +  G_{l_1 l_2}^{(2)} \enhUB_{l_1}(\tau_{13})\UF_{l_2}(\tau_{21}) +
	G_{l_1 l_2}^{(3)} \UF_{l_1}(\tau_{12})\enhUB_{l_2}(\tau_{23})
	\Big\},\label{eq:three-point-exp}
\end{align}
where $\enhUB_{l}(\tau)$ is an extended bosonic basis with two additional basis functions
$\set{   \frac{1}{\sqrt{\beta}} , \sqrt{\frac{3}{\beta}}(2\tau/\beta-1) }$.
Namely,
\begin{align}
	\enhUB_l(\tau) =\begin{cases}
	\frac{1}{\sqrt{\beta}} & (l=0)\\
	\sqrt{\frac{3}{\beta}}(2\tau/\beta-1) & (l=1) \\
	\UB_{l-2}(\tau) & (l \geq2)
	\end{cases}
\end{align}
and $\enhUB_l(\tau)$ are defined correspondingly.\footnote{While this makes the bosonic basis overcomplete, it does not cause any further difficulty in numerical calculations since the expansion formula for $\Gthree$ is already overcomplete.}.
We will see that $\enhUB_1(\tau)$ play an important role in expanding $\Gfour$. 
For a consistent description of $\Gthree$ and $\Gfour$,
we keep  $\enhUB_1(\tau)$  in Eq.~(\ref{eq:three-point-exp}).
Note that $\enhUB$ is a non-orthogonal basis set.

In the Matsubara domain, $\Gthree$ can be represented as
\begin{align}
&\Gthree(i\omega_1, i\omega_2)\nonumber\\
&\equiv \int_0^\beta d\tau_{13} d\tau_{23} e^{i\omega_1 \tau_{13} + i \omega_2 \tau_{23}} \Gthree(\tau_1, \tau_2, \tau_3)
\nonumber\\
& =  \sum_{l_1,l_2=0}^\infty \Big\{G^{(1)}_{l_1l_2}\UF_{l_1}(i\omega_1)\UF_{l_2}(i\omega_2) \nonumber\\
& \quad+  G_{l_1 l_2}^{(2)} \enhUB_{l_1}(i\omega_1 + i \omega_2)\UF_{l_2}(i \omega_2)
\nonumber\\
& \quad+
G_{l_1 l_2}^{(3)} \UF_{l_1}(i\omega_1)\enhUB_{l_2}(i\omega_1 + i \omega_2)
\Big\}.\label{eq:three-point-exp2}
\end{align}
The definition of the Fourier transform is given in Eq.~(\ref{eq:three-point-Fourier}).

As in Eq.~(\ref{eq:gl}), we can relate $G_{l_1 l_2}^{(n)}$ to the spectral functions $\rho^{(n)}(\epsilon_1, \epsilon_2)$. 
If $\wmax$ is large enough that $\rho^{(n)}(\epsilon_1, \epsilon_2)$ is bounded in [$-\wmax$, $\wmax$], we obtain
\begin{align}
	G_{l_1 l_2}^{(n)} &\propto s_{l_1}^{\alpha} s_{l_2}^{\alpha'} \rho^{(n)}_{l_1l_2},\label{eq:three-point-gl}
\end{align}
where $\alpha$ and $\alpha'$ are either F or $\overline{\mathrm{B}}$ depending on $n$, and
\begin{align}
	\rho^{(n)}_{l_1l_2} &\equiv \int_{-\wmax}^\wmax d \omega_1 d \omega_2~V^\alpha_{l_1}(\omega_1)V^{\alpha^\prime}_{l_2}(\omega_2) \rho^{(n)}(\omega_1, \omega_2).\label{eq:three-point-exp3}
\end{align}
Therefore, $G_{l_1 l_2}^{(n)}$ decays exponentially as in the case of the single-particle Green's function.

We propose to use Eqs.~(\ref{eq:three-point-exp}) and (\ref{eq:three-point-exp2}) as a compact representation for arbitrary $\Gthree$.
We note that this is an \emph{overcomplete and non-orthogonal representation}.
The overcomplete nature can be understood by considering the fact that each of the three terms in Eq.~(\ref{eq:three-point-exp}) is an expansion in terms of a complete basis.
The coefficients $G_{l_1 l_2}^{(n)}$, therefore, are not uniquely determined.
In other words, there is no inversion formula of Eqs.~(\ref{eq:three-point-exp}) and (\ref{eq:three-point-exp2}).
We will present one possible way to determine $G_{l_1 l_2}^{(n)}$ in Section~\ref{sec:coefficients}.

\subsection{The four-point Green's function}\label{sec:Gfour}
\begin{figure}
	\centering
	\includegraphics[width=0.3\textwidth,clip]{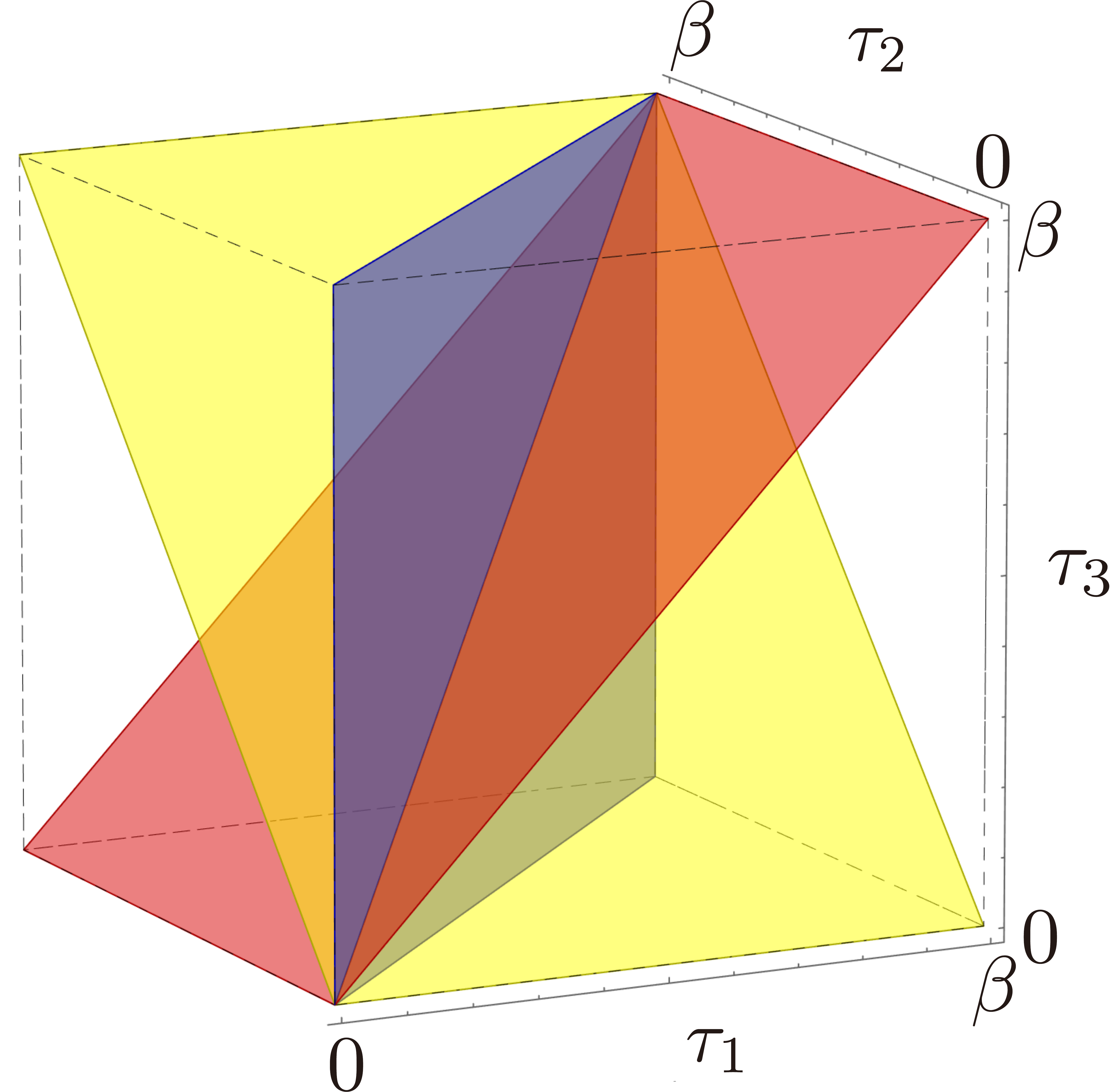}
	\caption{
		(Color online) Equal-time planes of $\Gfour(\tau_1, \tau_2, \tau_3, \tau_4)$ running diagonally through the cubic space: $\tau_1$ = $\tau_2$ (blue), $\tau_2 = \tau_3$ (yellow), $\tau_1 = \tau_3$ (red). We take $\tau_4 = 0$. These planes meet in the body diagonal $\tau_1 = \tau_2 = \tau_3$. Three additional planes $\tau_1, \tau_2, \tau_3 = 0,\beta$ constitute the bounding box (dashed).
		Adapted from Ref.~\onlinecite{WallerbergerPhD}.
	}
	\label{fig:equal-planes}
\end{figure}

We now derive a compact representation of the four-point Green's function defined by
\begin{align}
	\Gfour(\tau_1, \tau_2, \tau_3, \tau_4)
	=\braket{T_\tau A(\tau_1) B(\tau_2) C(\tau_3) D(\tau_4)},
	\label{eq:four-point-def}
\end{align}
where $A$, $B$, $C$, $D$ are all fermionic operators.
The derivation of the expansion formula for $\Gfour$ proceeds along the lines of the last section.
Equal-time planes, where $\Gfour(\tau_1, \tau_2, \tau_3, \tau_4)$ has discontinuities, are illustrated in Fig.~\ref{fig:equal-planes}.

\begin{figure}
	\centering
	\includegraphics[width=0.5\textwidth,clip]{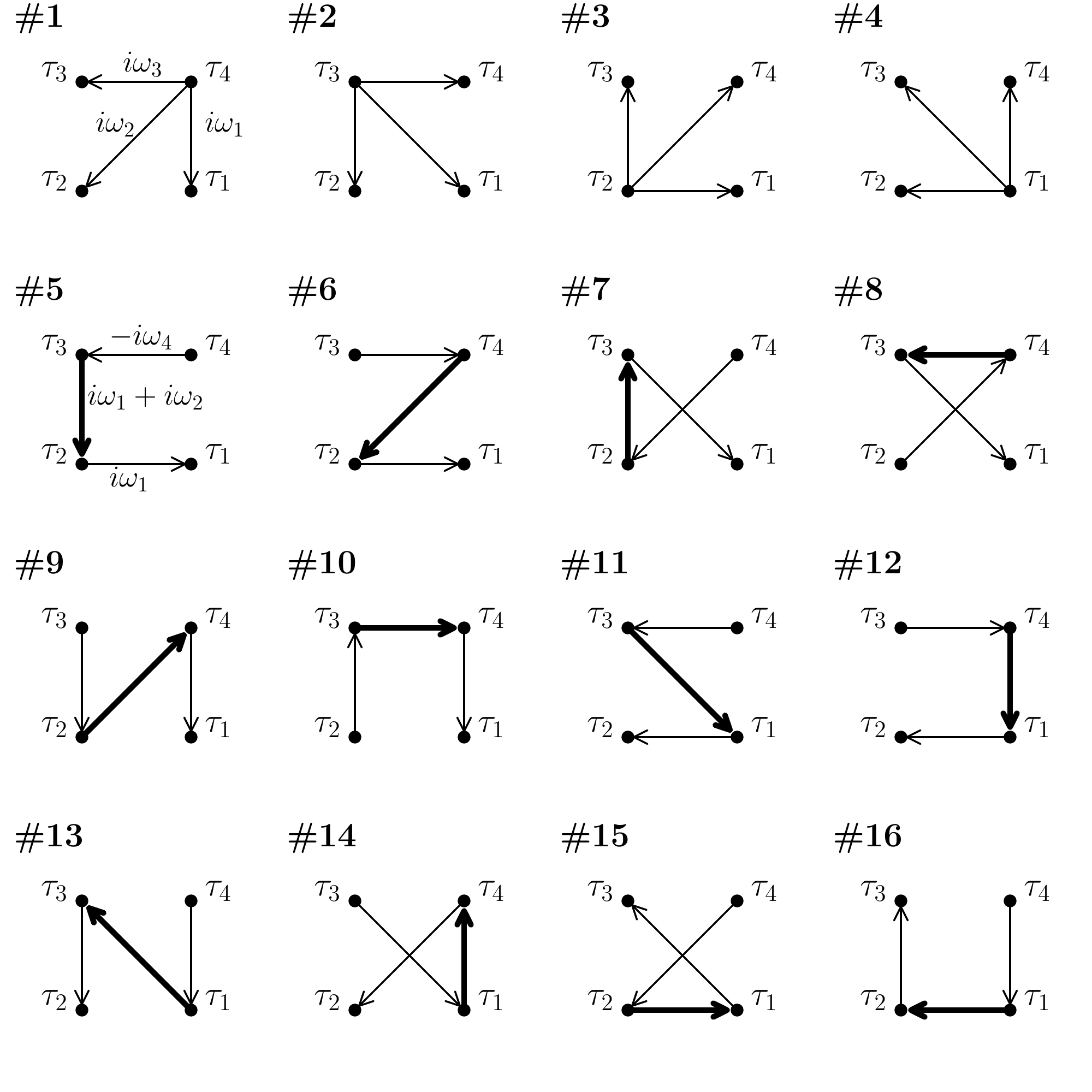}
	\caption{
		(Color online) Diagrams for the 16 representations for the four-point Green's function in Table~\ref{table:summary}.
		The bold arrows denote imaginary times with the bosonic statistics.
The incoming flux at each vertex $\tau_i$ equals to $i \omega_i$ as demonstrated for \# 1 and \# 5.
			This conservation law can be used for generating entries in Table.~\ref{table:summary}.
	}
	\label{fig:diagram}
\end{figure}

There are 16 distinct ways of defining three relative times, which are summarized in Table~\ref{table:summary}.
Figure~\ref{fig:diagram} illustrates how those combinations are generated.

\begin{table}
	\centering
	\begin{tabular}{c|c|c|c}
		\hline
	$\#r$ & ($i\omega$, $i\omega$', $i\omega$'') & ($\tau$, $\tau'$, $\tau''$) & ($\alpha$, $\alpha'$, $\alpha''$) \\
		\hline
		\hline
\#1 & ($i\omega_1$, $i\omega_2$, $i\omega_3$) & ($\tau_{14}$, $\tau_{24}$, $\tau_{34}$) &  (F,F,F)  \\
\#2 & ($i\omega_1$, $i\omega_2$, $i\omega_4$) & ($\tau_{13}$, $\tau_{23}$, $\tau_{43}$) &  (F,F,F)  \\
\#3 & ($i\omega_1$, $i\omega_3$, $i\omega_4$) & ($\tau_{12}$, $\tau_{32}$, $\tau_{42}$) &  (F,F,F)  \\
\#4 & ($i\omega_2$, $i\omega_3$, $i\omega_4$) & ($\tau_{21}$, $\tau_{31}$, $\tau_{41}$) &  (F,F,F)  \\ \hline
\#5 & ($i\omega_1$, $i\omega_1 + i\omega_2$, $-i\omega_4$) & ($\tau_{12}$, $\tau_{23}$, $\tau_{34}$) &  (F,$\overline{\mathrm{B}}$,F)  \\
\#6 & ($i\omega_1$, $i\omega_1 + i\omega_2$, $-i\omega_3$) & ($\tau_{12}$, $\tau_{24}$, $\tau_{43}$) &  (F,$\overline{\mathrm{B}}$,F)  \\
\#7 & ($i\omega_1$, $i\omega_1 + i\omega_3$, $-i\omega_4$) & ($\tau_{13}$, $\tau_{32}$, $\tau_{24}$) &  (F,$\overline{\mathrm{B}}$,F)  \\
\#8 & ($i\omega_1$, $i\omega_1 + i\omega_3$, $-i\omega_2$) & ($\tau_{13}$, $\tau_{34}$, $\tau_{42}$) &  (F,$\overline{\mathrm{B}}$,F)  \\
\#9 & ($i\omega_1$, $i\omega_1 + i\omega_4$, $-i\omega_3$) & ($\tau_{14}$, $\tau_{42}$, $\tau_{23}$) &  (F,$\overline{\mathrm{B}}$,F)  \\
\#10 & ($i\omega_1$, $i\omega_1 + i\omega_4$, $-i\omega_2$) & ($\tau_{14}$, $\tau_{43}$, $\tau_{32}$) &  (F,$\overline{\mathrm{B}}$,F)  \\
\#11 & ($i\omega_2$, $i\omega_2 + i\omega_1$, $-i\omega_4$) & ($\tau_{21}$, $\tau_{13}$, $\tau_{34}$) &  (F,$\overline{\mathrm{B}}$,F)  \\
\#12 & ($i\omega_2$, $i\omega_2 + i\omega_1$, $-i\omega_3$) & ($\tau_{21}$, $\tau_{14}$, $\tau_{43}$) &  (F,$\overline{\mathrm{B}}$,F)  \\
\#13 & ($i\omega_2$, $i\omega_2 + i\omega_3$, $-i\omega_4$) & ($\tau_{23}$, $\tau_{31}$, $\tau_{14}$) &  (F,$\overline{\mathrm{B}}$,F)  \\
\#14 & ($i\omega_2$, $i\omega_2 + i\omega_4$, $-i\omega_3$) & ($\tau_{24}$, $\tau_{41}$, $\tau_{13}$) &  (F,$\overline{\mathrm{B}}$,F)  \\
\#15 & ($i\omega_3$, $i\omega_3 + i\omega_1$, $-i\omega_4$) & ($\tau_{31}$, $\tau_{12}$, $\tau_{24}$) &  (F,$\overline{\mathrm{B}}$,F)  \\
\#16 & ($i\omega_3$, $i\omega_3 + i\omega_2$, $-i\omega_4$) & ($\tau_{32}$, $\tau_{21}$, $\tau_{14}$) &  (F,$\overline{\mathrm{B}}$,F)  \\
		\hline
	\end{tabular}
	\caption{16 different notations of relative times for $\Gfour(\tau_1,\tau_2,\tau_3,\tau_4)$, the corresponding Matsubara frequencies and statistics.}
	\label{table:summary}
\end{table}

The spectral representation of $\Gfour(\tau_1, \tau_2, \tau_3, \tau_4)$ was derived in Ref.~\onlinecite{Hafermann09b} with explicit consideration of singular contributions.
Analyzing this spectral representation,
we obtain an expansion formula consisting of $16$ representations:
\begin{align}
	&\Gfour(\tau_1, \tau_2, \tau_3, \tau_4)
	\nonumber\\
	& = \sum_{l_1, l_2, l_3=0}^{\infty} \Big\{
	G_{l_1 l_2 l_3}^{(1)}  \UF_{l_1}(\tau_{14}) \UF_{l_2}(\tau_{24}) \UF_{l_3}(\tau_{34})
	\nonumber\\
	&\hspace{4em}+ \cdots \nonumber\\
	&\hspace{4em}+ G_{l_1 l_2 l_3}^{(16)} \UF_{l_1}(\tau_{32}) \enhUB_{l_2}(\tau_{21}) \UF_{l_3}(\tau_{14})
	\Big\}
	\nonumber\\
	& \equiv \sum_{r=1}^{16} \sum_{l_1, l_2, l_3=0}^{\infty} G_{l_1 l_2 l_3}^{(r)} U^{\alpha}_{l_1}(\tau) U^{\alpha'}_{l_2}(\tau') U^{\alpha''}_{l_3}(\tau''),
	\label{eq:ir-four-point}
\end{align}
where $(\alpha, \alpha', \alpha'')$ and ($\tau$, $\tau'$, $\tau''$) depend on $r$ according to Table~\ref{table:summary}.
Here, we introduced $\enhUB_l(\tau^\prime)$ ($l=0,1$) to describe the singular contribution systematically.
See Appendix~\ref{appendix:four-point} for the derivation and detailed discussion.

The corresponding Matsubara representation reads
\begin{align}
	& \Gfour(i\omega_1, i\omega_2, i \omega_3, i \omega_4 ) \nonumber \\
	& \equiv \beta^{-1}\int_0^\beta d\tau_1 d\tau_2 d\tau_3 d\tau_4 e^{i(\omega_1 \tau_1 +  \omega_2 \tau_2 +  \omega_3 \tau_3 + \omega_4 \tau_4)}  \nonumber\\
	&\hspace{1em} \times\Gfour(\tau_1, \tau_2, \tau_3, \tau_4)\nonumber\\
	& = \delta_{\omega_1 + \omega_2 + \omega_3 + \omega_4,0}
	\nonumber\\
	&\times \sum_{l_1, l_2, l_3=0}^{\infty} \Big\{
	G_{l_1 l_2 l_3}^{(1)}  \UF_{l_1}(i\omega_1) \UF_{l_2}(i\omega_2) \UF_{l_3}(i\omega_3)
	\nonumber\\
	&\hspace{4em}+ \cdots
	\nonumber\\
	&\hspace{4em}+ G_{l_1 l_2 l_3}^{(16)} \UF_{l_1}(i \omega_3) \enhUB_{l_2}(i \omega_3 + i \omega_2) \UF_{l_3}(-i\omega_4) \Big\}\nonumber \\
	& \equiv \delta_{\omega_1 + \omega_2 + \omega_3 + \omega_4,0}
	\nonumber\\
	&\times \sum_{r=1}^{16} \sum_{l_1 l_2 l_3} G_{l_1 l_2 l_3}^{(r)} U^{\alpha}_{l_1}(i\omega) U^{\alpha'}_{l_2}(i\omega') U^{\alpha''}_{l_3}(i\omega''),
	\label{eq:ir-four-point2}
\end{align}
where $(i\omega, i\omega', i\omega'')$ as well as $(\alpha, \alpha', \alpha'')$ depend on $r$ as listed in Table~\ref{table:summary}.

As detailed in Appendix.~\ref{appendix:four-point},
one can show that the expansion coefficients decay as
\begin{align}
	G^{(r)}_{l_1 l_2 l_3}  &\propto s_{l_1}^{\alpha} s_{l_2}^{\alpha'} s_{l_3}^{\alpha''} \rho_{l_1 l_2 l_3}^{(r)},\label{eq:four-point-decay}
\end{align}
when $\wmax$ is chosen to be sufficiently large.%

\subsection{Systems with multiple degrees of freedoms}
The current expression formula can therefore be extended to systems with multiple degrees of freedom such as a multi-orbital systems. 
As an illustration, let us consider the four-point Green's function for a multi-orbital system:
\begin{align}
\Gfour_{abcd}(\tau_1, \tau_2, \tau_3, \tau_4) &= \braket{T_\tau c_a(\tau_1) c^\dagger_b(\tau_2) c_c(\tau_3) c^\dagger_d(\tau_4)},
\end{align}
where $a$, $b$, $c$, $d$ are the combined indices of spin and orbital.
The expression formula for this Green's function reads
\begin{align}
&\Gfour_{abcd}(\tau_1, \tau_2, \tau_3, \tau_4)
\nonumber\\
& \equiv \sum_{r=1}^{16} \sum_{l_1, l_2, l_3=0}^{\infty} G_{a b c d; l_1 l_2 l_3}^{(r)} U^{\alpha}_{l_1}(\tau) U^{\alpha'}_{l_2}(\tau') U^{\alpha''}_{l_3}(\tau'').
\end{align}
One can see that the decomposition into 16 parts applies to each orbital pair $(a,b,c,d)$ .

\section{Accuracy of the compact representation of two-particle Green's functions}\label{sec:fit}
\label{sec:coefficients}
The expansion formulas presented in the previous section are based on an overcomplete basis set.
Hence, the expansion coefficients are not uniquely determined for given imaginary-time data.
In this section, we present one possible way to determine these coefficients, and demonstrate accuracy and compactness of our representation.

As a simple example,
we consider a single-orbital single-site Hubbard model 
whose Hamiltonian is defined as
\begin{align}
	\mathcal{H} &= U n_{\uparrow} n_{\downarrow} - \mu ( n_{\uparrow} +n_{\downarrow} ).\label{eq:Hubbard-atom}
\end{align}
We solve this model for $U=2$, $\mu=U/2$ and $\beta=20$ using exact diagonalization.

\subsection{Three-point Green's function}
\label{sec:coefficients-3pt}
\begin{figure}
	\centering
	\includegraphics[width=0.45\textwidth,clip]{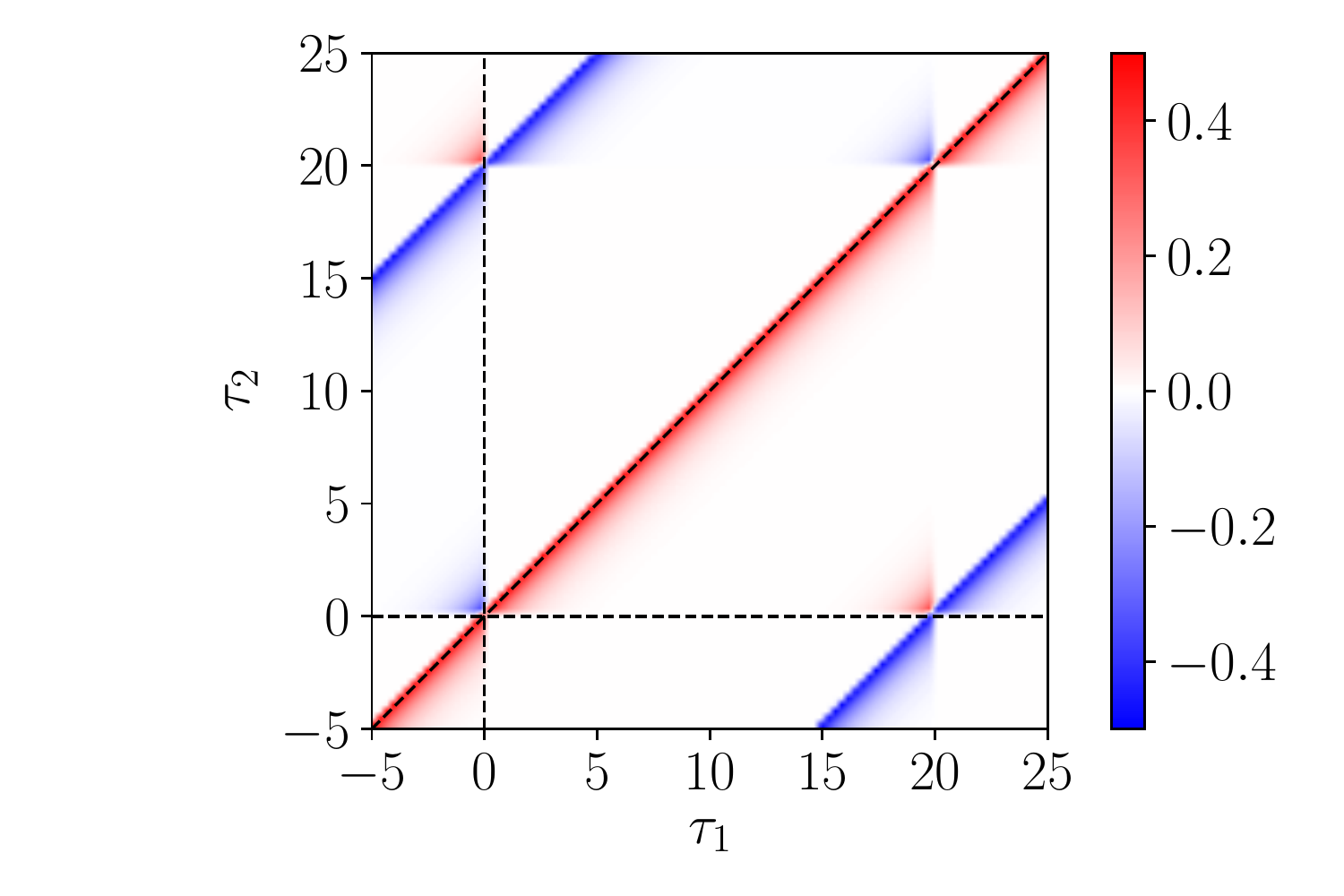}
	\caption{
		(Color online) $\Gthree(\tau_1, \tau_2, 0)$ for the single-site Hubbard model obtained from an exact diagonalization of the system.
	}
	\label{fig:sampling-point}
\end{figure}

We first consider the three-point Green's function defined by
\begin{align}
	\Gthree(\tau_1, \tau_2, 0) &= \braket{T_\tau c_{\uparrow}(\tau_1) c_{\uparrow}^\dagger (\tau_2) n_{\uparrow}(0)}.
\end{align}
The analytic expression of $\Gthree(\tau_1, \tau_2, \tau_3)$ is given in Appendix~\ref{appendix:single-site}.
We plot $\Gthree(\tau_1, \tau_2, \tau_3)$ in Fig.~\ref{fig:sampling-point}.
There are discontinuities at $\tau_1 = \tau_2$, $\tau_1=n\beta$ and $\tau_2=n\beta$ ($n=0,\pm 1, \pm 2,\cdots$).

We expand $\Gthree(\tau_1, \tau_2, 0)$ in the form of Eq.~(\ref{eq:three-point-exp}) 
using an IR basis $\{ U^{\alpha}_l (\tau) \}$  constructed for $\beta\wmax=40$.
Basis functions corresponding to small singular values $s_l^{\alpha}$ are irrelevant [see Eq.~(\ref{eq:three-point-gl})].
Therefore, we truncate small singular values $s_l^\mathrm{F}/s_0^\mathrm{F} \lesssim 2.5 \times 10^{-6}$ and
$s_l^{\overline{\mathrm{B}}}/s^{\overline{\mathrm{B}}}_0 \lesssim 3.5 \times 10^{-5}$.
As a result, we keep $N_\mathrm{IR}=16$ basis functions for fermions ($\alpha=\mathrm{F}$) and bosons ($\alpha=\overline{\mathrm{B}}$), respectively.
This leaves $3N_\mathrm{IR}^2$ coefficients of $G_{l_1 l_2}^{(r)}$ to be determined.

We now discuss how to obtain the basis expansion coefficients for a given imaginary-time object.
 We found it more convenient to fit imaginary-time data on a uniform grid rather than to perform a basis transformation.
This is because the latter one would require the computation of products of the basis vectors between the imaginary-time object and involve numerical integration in two dimensions.

With the conditions above, Eq.~(\ref{eq:three-point-exp}) may be rewritten symbolically as
\begin{align}
\boldsymbol{G}^\mathrm{3pt} = \boldsymbol{A} \boldsymbol{G}^\mathrm{IR},
\label{eq:three-point-exp-matrix}
\end{align}
where the vector $\boldsymbol{G}^\mathrm{3pt}$ is a one-dimensional expression for $\Gthree(\tau_1, \tau_2, 0)$ computed on $N_\mathrm{smp}=50^2$ uniform grid points, and $\boldsymbol{G}^\mathrm{IR}$ for $G_{l_1 l_2}^{(r)}$.
$\boldsymbol{A}$ is a matrix of size $(N_\mathrm{smp} \times 3N_\mathrm{IR}^2)$ that stores coefficients in Eq.~(\ref{eq:three-point-exp}).
The simplest way to invert Eq.~(\ref{eq:three-point-exp-matrix}) is the least square fitting. However, this fitting procedure suffers from a numerical instability due to redundant degrees of freedom in the overcomplete representation.

To avoid this instability, we use the so-called Ridge regression whose cost function is given by
\begin{align}
\mathcal{L} &= \| \boldsymbol{G}^\mathrm{3pt} - \boldsymbol{A} \boldsymbol{G}^\mathrm{IR} \|^2
+ \lambda \sum_{r=1}^{3} \sum_{l_1, l_2=0}^{N_\mathrm{IR}-1} \mid  G^{(r)}_{l_1 l_2}/S^{(r)}_{l_1 l_2} \mid^2.\label{eq:Ridge-Gthree}
\end{align}
Here, the first term denotes the ordinary Euclidean norm of the residual vector.
We defined $S_{l_1 l_2}^{(r)}\equiv s^\alpha_{l_1} s^{\alpha^\prime}_{l_2}$ [see Eq.~(\ref{eq:three-point-gl})]
with ``singular values" for $\enhUB$ being
\begin{align}
s^\overlineB_l \equiv \begin{cases}
s^\mathrm{B}_0 & (l=0,1)\\
s^\mathrm{B}_{l-2} & (l \geq2)
\end{cases}.
\end{align}
The second term makes a difference to the least square method:
A solution having a small norm (weighted by $S_{l_1 l_2}^{(r)}$) is selected out of many degenerate solutions for the least-squares fit due to the overcompleteness.~\footnote{In the study of analytical continuation in Ref.~\onlinecite{Otsuki17}, the $L_1$-norm minimizing solution is considered instead to eliminate irrelevant coefficients. In the present case, the $L_2$-norm minimization is sufficient since we are interested in the fitting of the imaginary-time data but not in the spectrum.}
Thus, we take $\lambda = 10^{-10}$, which is a small value larger than machine precision.
Appendix~\ref{appendix:ridge} contains an explicit expression for the solution of the Ridge regression.

\begin{figure}
	\centering
	\includegraphics[width=0.4\textwidth,clip]{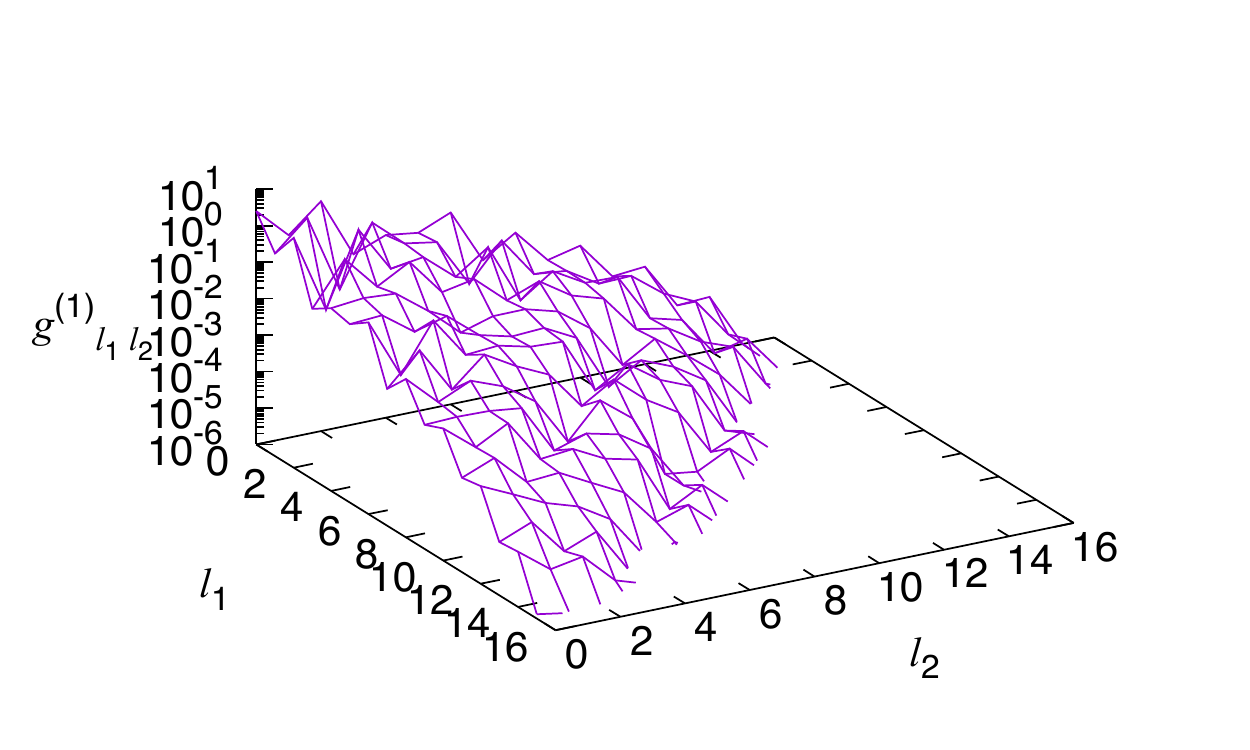}
	\includegraphics[width=0.4\textwidth,clip]{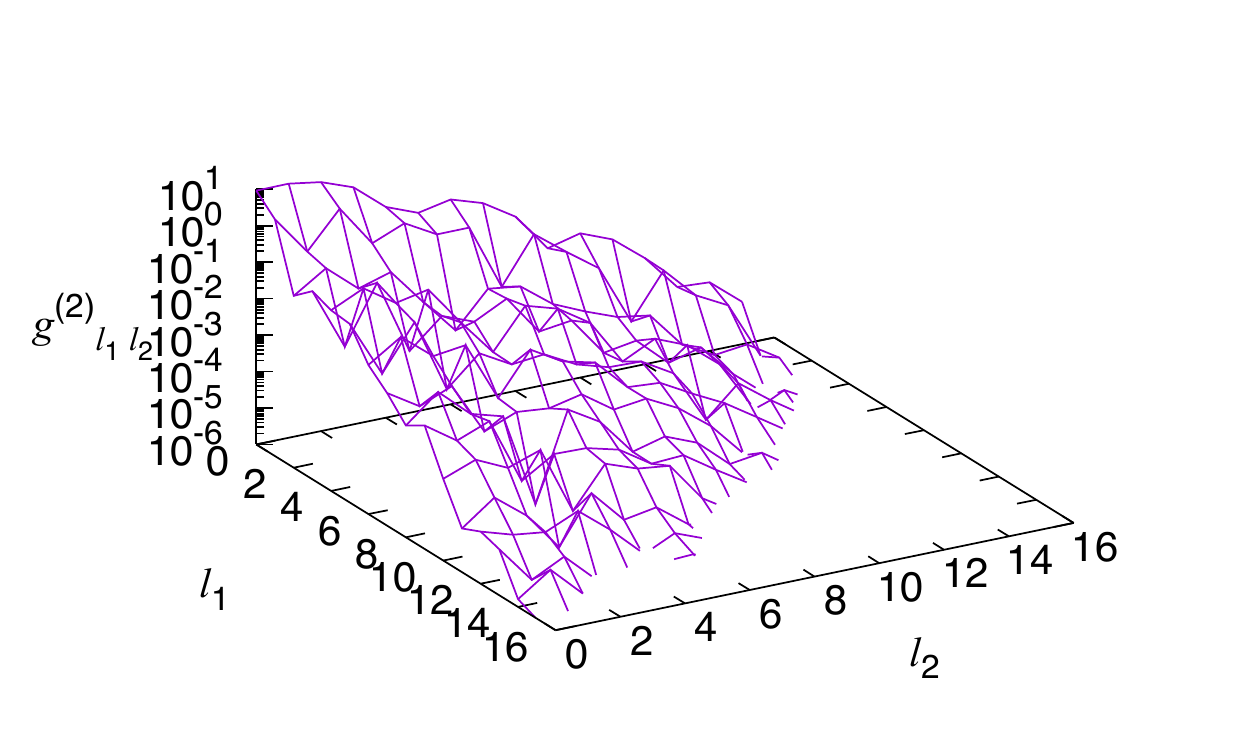}
	\includegraphics[width=0.4\textwidth,clip]{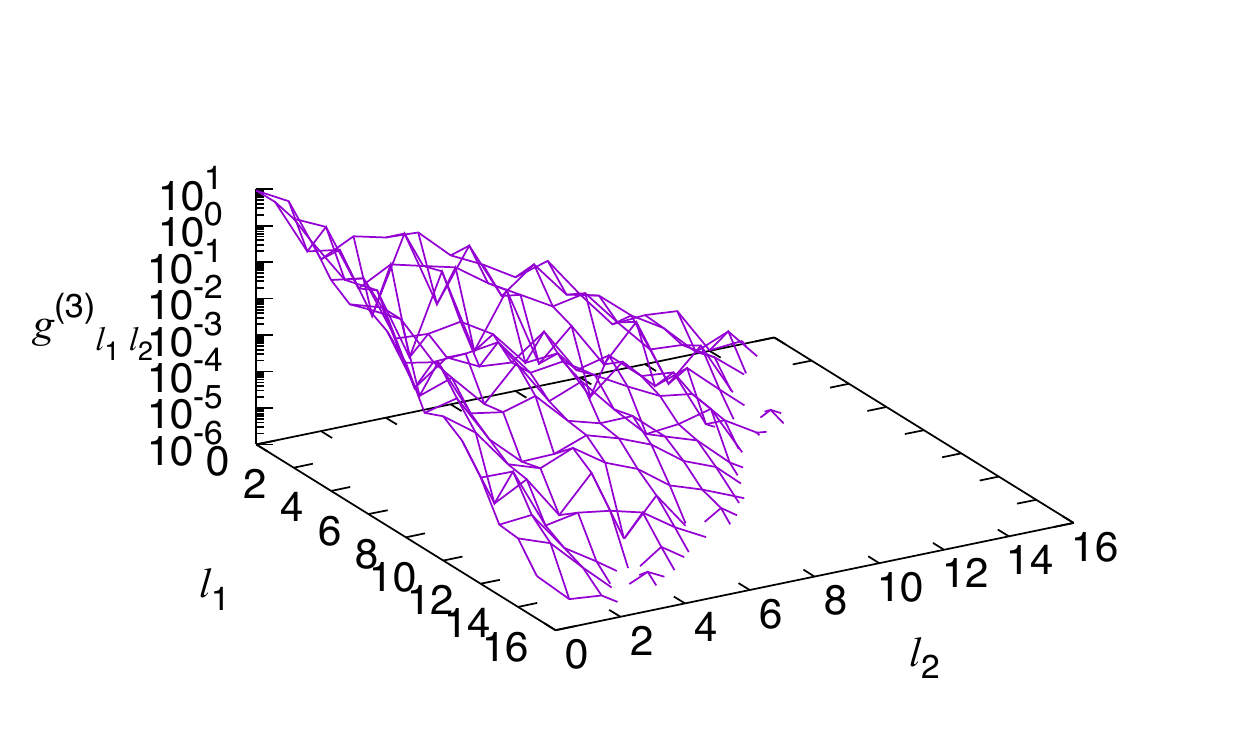}
	\includegraphics[width=0.4\textwidth,clip]{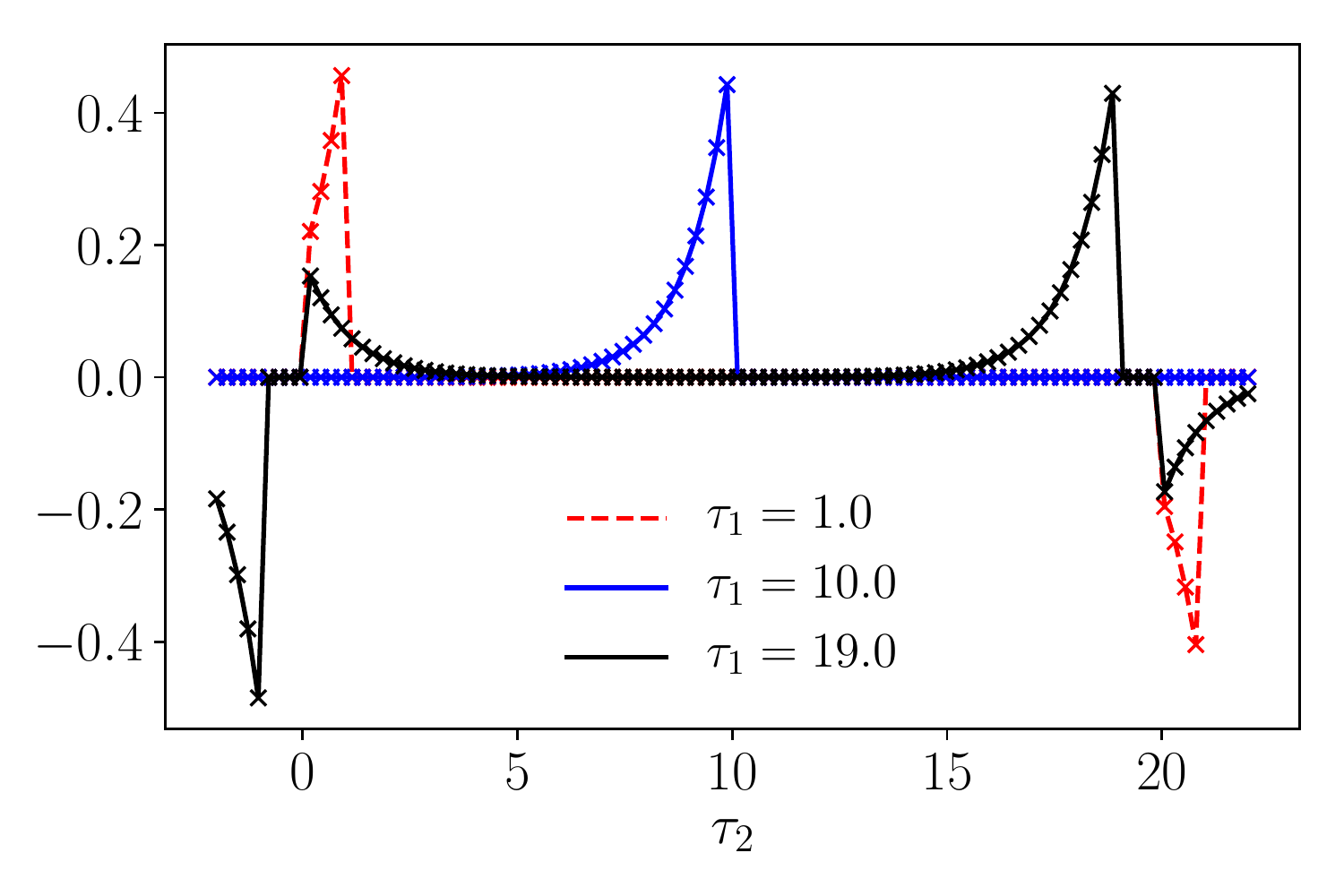}
	
	\caption{
		(Color online) Upper three panels: parameters obtained for the three-point Green's function.
		Lower panel: comparison of exact values and interpolated ones.
	}
	\label{fig:three-point-coeff}
\end{figure}

Figure~\ref{fig:three-point-coeff}(a) shows the results for the expansion coefficients $G_{l_1 l_2}^{(r)}$ obtained in the way explained above.
As expected, the coefficients decay exponentially.
Figure~\ref{fig:three-point-coeff}(b) shows a comparison between the exact data of $\Gthree(\tau_1, \tau_2, 0)$ and data evaluated from $G_{l_1 l_2}^{(r)}$ using Eq.~(\ref{eq:three-point-exp}).
The exact data is correctly reproduced and the solution includes an accurate description of all discontinuities,
demonstrating that this compact representation is accurate to within the tolerance given by the singular-value cutoff.

\subsection{Four-point Green's function}
\begin{figure}
	\centering
	\includegraphics[width=0.45\textwidth,clip]{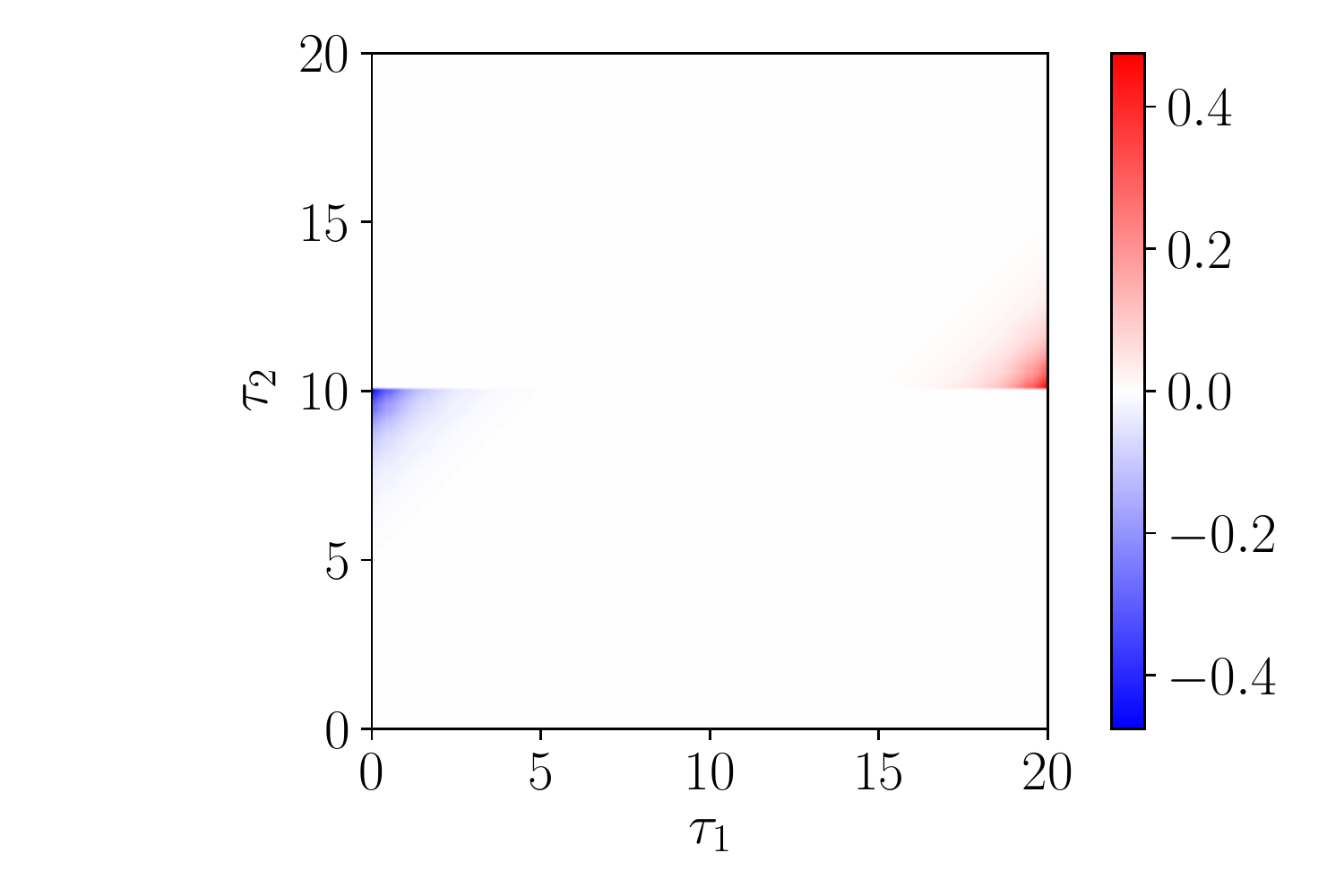}
	\caption{
		(Color online) $\Gfour(\tau_1, \tau_2, \tau_3, \tau_4)$ in Eq.~(\ref{eq:four-point-example}) computed by the exact diagonalization at $\tau_3=\beta/2$ and $\tau_4=0$.
	}
	\label{fig:four-point-map}
\end{figure}

We now test our scheme for the four-point Green's function defined by
\begin{align}
	\Gfour(\tau_1, \tau_2, \tau_3, \tau_4) &= \braket{T_\tau c_\uparrow(\tau_1) c^\dagger_\uparrow(\tau_2) c_\uparrow(\tau_3) c^\dagger_\uparrow(\tau_4)}.
\label{eq:four-point-example}
\end{align}
Figure~\ref{fig:four-point-map} shows $\Gfour$ computed at $\tau_3=\beta/2$ and $\tau_4=0$.

Since $\Gfour$ now includes the same operators ($A=C$, $B=D$), we can reduce the number of expansion coefficients using the crossing symmetries,
e.g., $\Gfour(\tau_1, \tau_2, \tau_3, \tau_4) = - \Gfour(\tau_3, \tau_2, \tau_1, \tau_4)$,
The 16 representations in Eq.~(\ref{eq:ir-four-point}) are classified into 6 subgroups under symmetry operations:
$(\#1, \#3)$,
$(\#2, \#4)$,
$(\#5, \#10, \#12, \#16)$,
$(\#6, \#9)$,
$(\#7, \#8, \#14, \#15)$,
$(\#11, \#13)$.
The coefficients $G^{(r)}_{l_1 l_2 l_3}$ in each subgroup are connected with each other.
Thus, Eq.~(\ref{eq:ir-four-point}) is reduced to
\begin{align}
	& \Gfour(\tau_1, \tau_2, \tau_3, \tau_4) \nonumber \\
	& = \hat{C} \frac{1}{2} \sum_{l_1 l_2 l_3} \Big\{
	G^{(1)}_{l_1 l_2 l_3} \UF_{l_1}(\tau_{14}) \UF_{l_2}(\tau_{24}) \UF_{l_3}(\tau_{34})\nonumber\\
	&\hspace{5em}+ G^{(2)}_{l_1 l_2 l_3} \UF_{l_1}(\tau_{13}) \UF_{l_2}(\tau_{23}) \UF_{l_3}(\tau_{43})\nonumber\\
	&\hspace{5em}+ 2G^{(5)}_{l_1 l_2 l_3} \UF_{l_1}(\tau_{12}) \enhUB_{l_2}(\tau_{23}) \UF_{l_3}(\tau_{34})\nonumber\\
	&\hspace{5em}+ G^{(6)}_{l_1 l_2 l_3} \UF_{l_1}(\tau_{12}) \enhUB_{l_2}(\tau_{24}) \UF_{l_3}(\tau_{43})\nonumber\\
	&\hspace{5em}+ 2G^{(7)}_{l_1 l_2 l_3} \UF_{l_1}(\tau_{13}) \enhUB_{l_2}(\tau_{32}) \UF_{l_3}(\tau_{24})\nonumber\\
	&\hspace{5em}+ G^{(11)}_{l_1 l_2 l_3} \UF_{l_1}(\tau_{21}) \enhUB_{l_2}(\tau_{13}) \UF_{l_3}(\tau_{34})
	\Big\},\label{eq:four-point-fit-model}
\end{align}
where the crossing-symmetry operator $\hat{C}$ acts on a function $f(\tau_1, \tau_2, \tau_3, \tau_4)$ as
\begin{align}
	\hat{C} f(\tau_1, \tau_2, \tau_3, \tau_4) &= f(\tau_1, \tau_2, \tau_3, \tau_4) - f(\tau_3, \tau_2, \tau_1, \tau_4) \nonumber\\
	& - f(\tau_1, \tau_4, \tau_3, \tau_2)  + f(\tau_3, \tau_4, \tau_1, \tau_2).
\end{align}
Using the same IR basis as in Sec.~\ref{sec:coefficients-3pt}, we have $6N_\mathrm{IR}^3$ expansion parameters to be fitted.

\begin{figure}
	\centering
	\includegraphics[width=0.4\textwidth,clip]{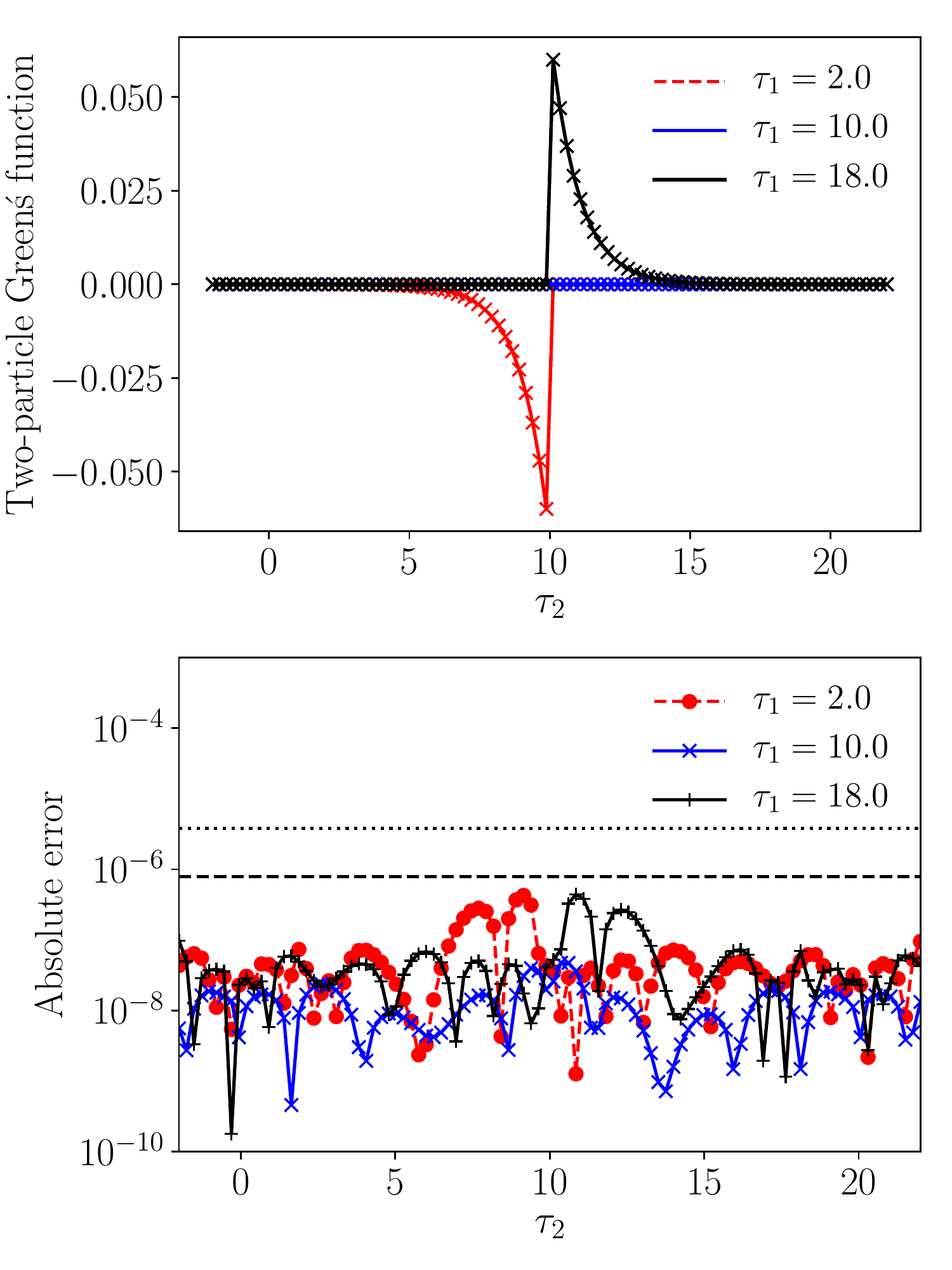}
	\caption{
		(Color online) (a) Comparison between the exact values of $\Gfour(\tau_1, \tau_2, \tau_3, \tau_4)$ and those evaluated by our compact representation at $\tau_3=\beta/2$ and $\tau_4=0$. (b) Absolute errors between the exact values and the interpolated data. The broken and dotted lines denote the cut-off values for singular values $s_l^\mathrm{F}/s_0^\mathrm{F} \simeq 2.5 \times 10^{-6}$ and $s_l^{\overline{\mathrm{B}}}/s^{\overline{\mathrm{B}}}_0 \simeq 3.5 \times 10^{-5}$, respectively.
	}
	\label{fig:four-point-cut}
\end{figure}

Imaginary-time data of $\Gfour(\tau_1, \tau_2, \tau_3, \tau_4)$ were prepared on non-uniform grids of $N_\mathrm{smp}=3\times 16^3$ points which represents the complicated $\tau$'s dependence efficiently.
We refer the interested reader to Appendix~\ref{appendix:grid} for more details.
As in the case of the three-point Green's function,
we computed the expansion parameters by means of the Ridge regression with $\lambda=10^{-8}$.
Figures~\ref{fig:four-point-cut}(a) and \ref{fig:four-point-cut}(b) show the result of the fitting of $\Gfour(\tau_1, \tau_2, \tau_3, \tau_4)$ and its absolute errors, respectively.
The errors are of the order of $10^{-6}$, being consistent with the cutoff in the singular values.

\section{Analysis of quantum Monte Carlo data for the Hubbard model}\label{sec:qmc}
\begin{figure}
	\begin{flushleft}\hspace{4em}(a) Real part \end{flushleft}
	\centering
	\includegraphics[width=0.4\textwidth,clip]{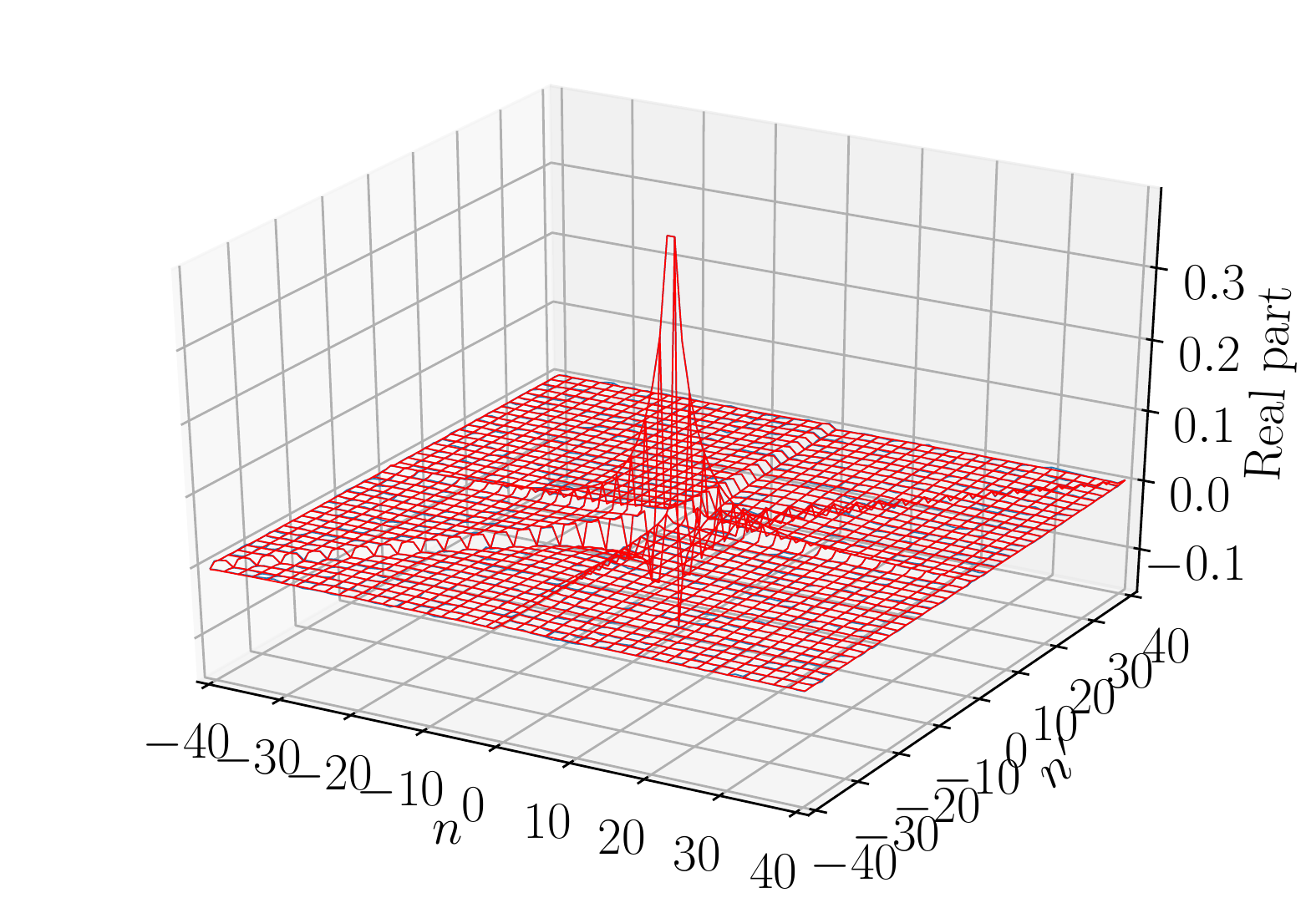}
	\includegraphics[width=0.35\textwidth,clip]{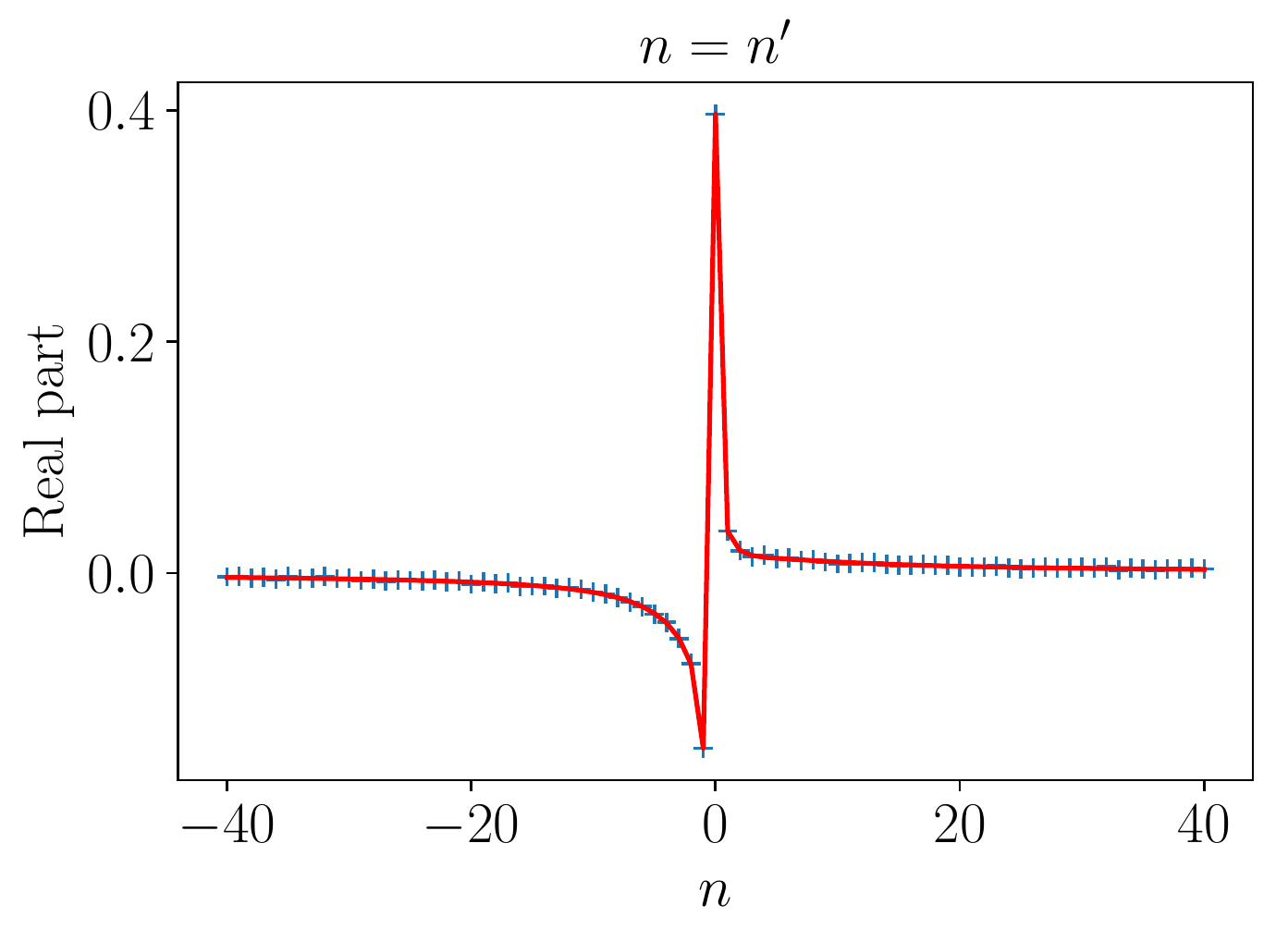}
	\begin{flushleft}\hspace{4em}(b) Imaginary part \end{flushleft}
	\centering	
	\includegraphics[width=0.4\textwidth,clip]{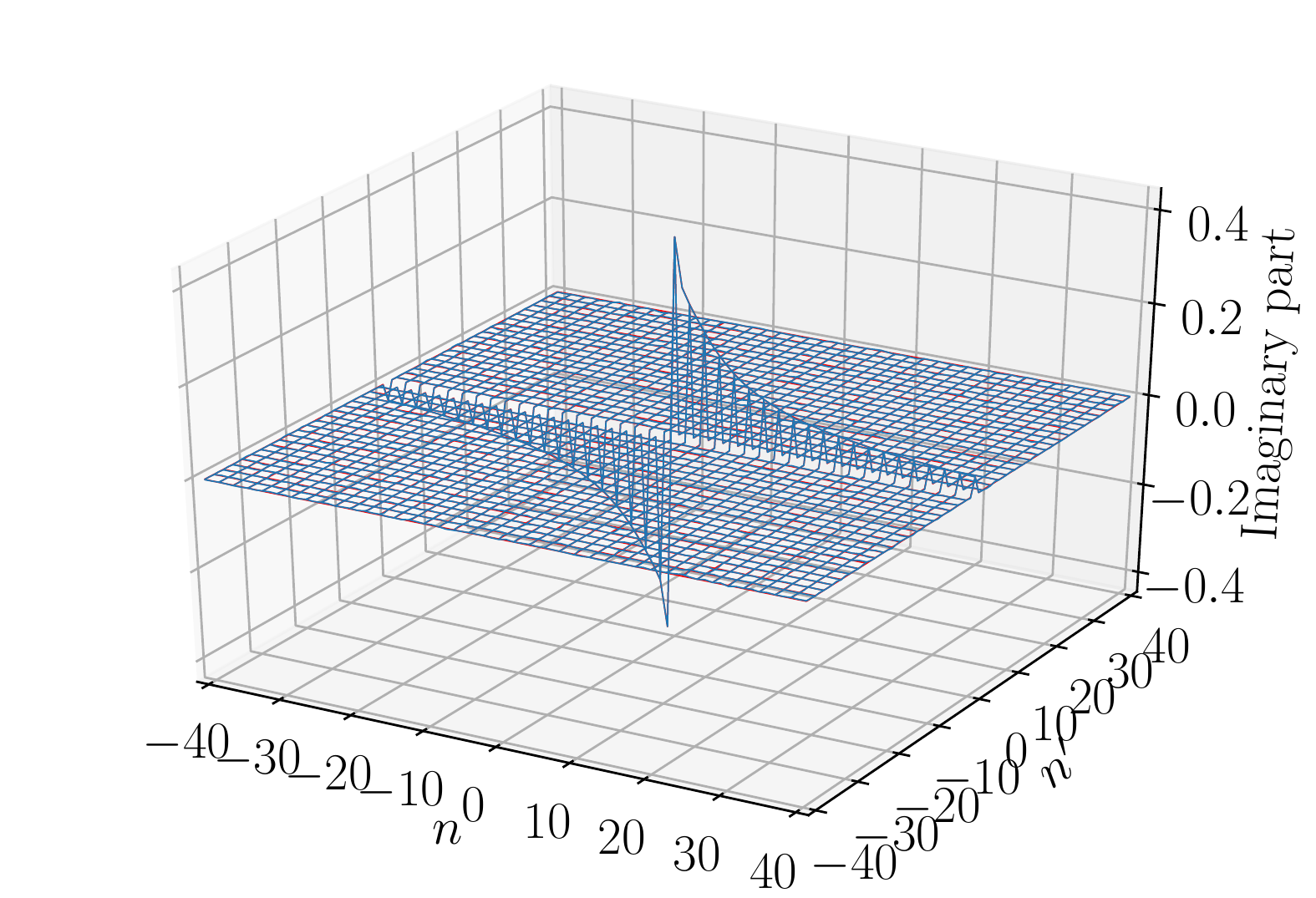}
	\includegraphics[width=0.35\textwidth,clip]{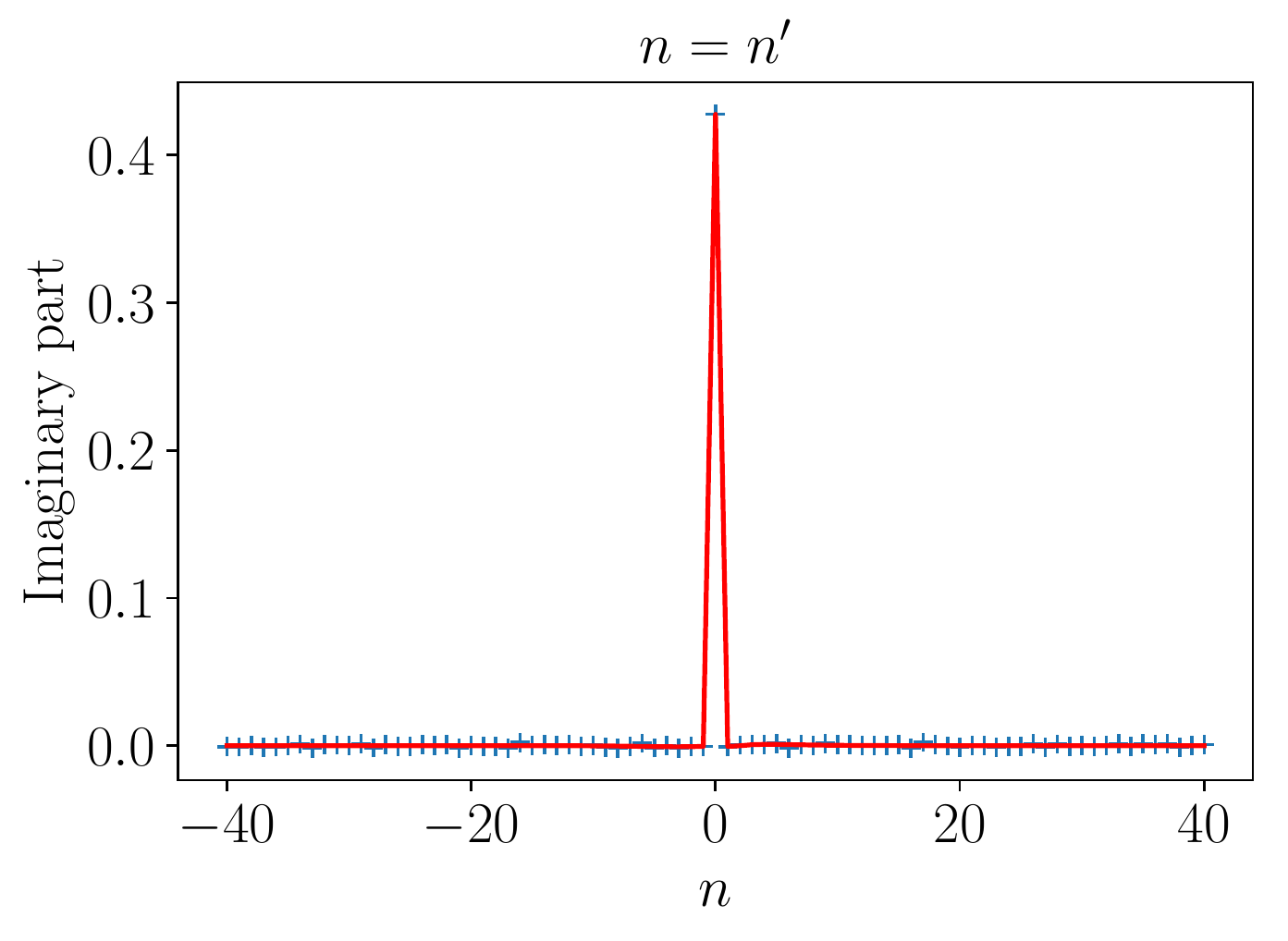}
	\caption{
		(Color online) Comparison of QMC data and the fit by the expansion formula in Eq.~(\ref{eq:three-point-exp2}).
		In the 3D plots, the QMC data and the fit are denoted by the blue and red lines, respectively.
		They are almost on the top of each other.
		In the 2D plots, the QMC data and the fit are denoted by the blue crosses and the red lines, respectively.
		In the all panels,
		we show the data only for $-40 \leq n \leq 40$ and $-40 \leq n^\prime \leq 40$.
	}
	\label{fig:qmc}
\end{figure}
\begin{figure}
	\centering
	\includegraphics[width=0.4\textwidth,clip]{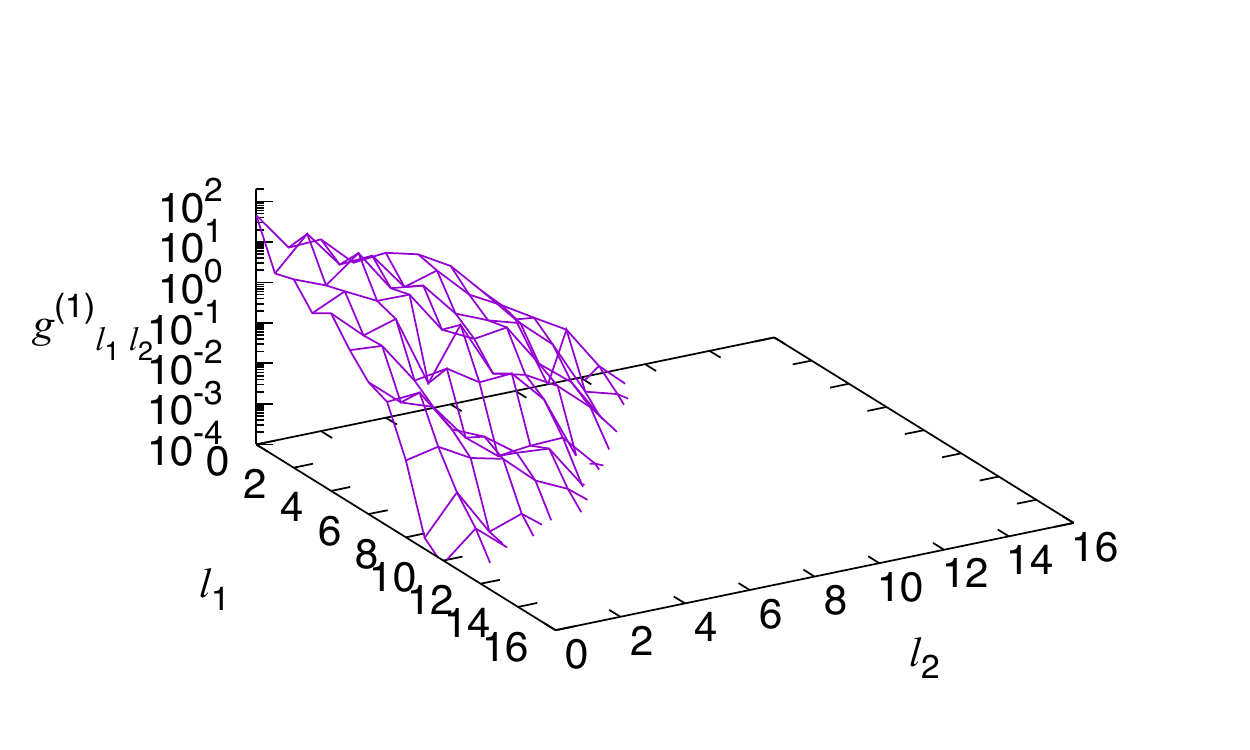}
	\includegraphics[width=0.4\textwidth,clip]{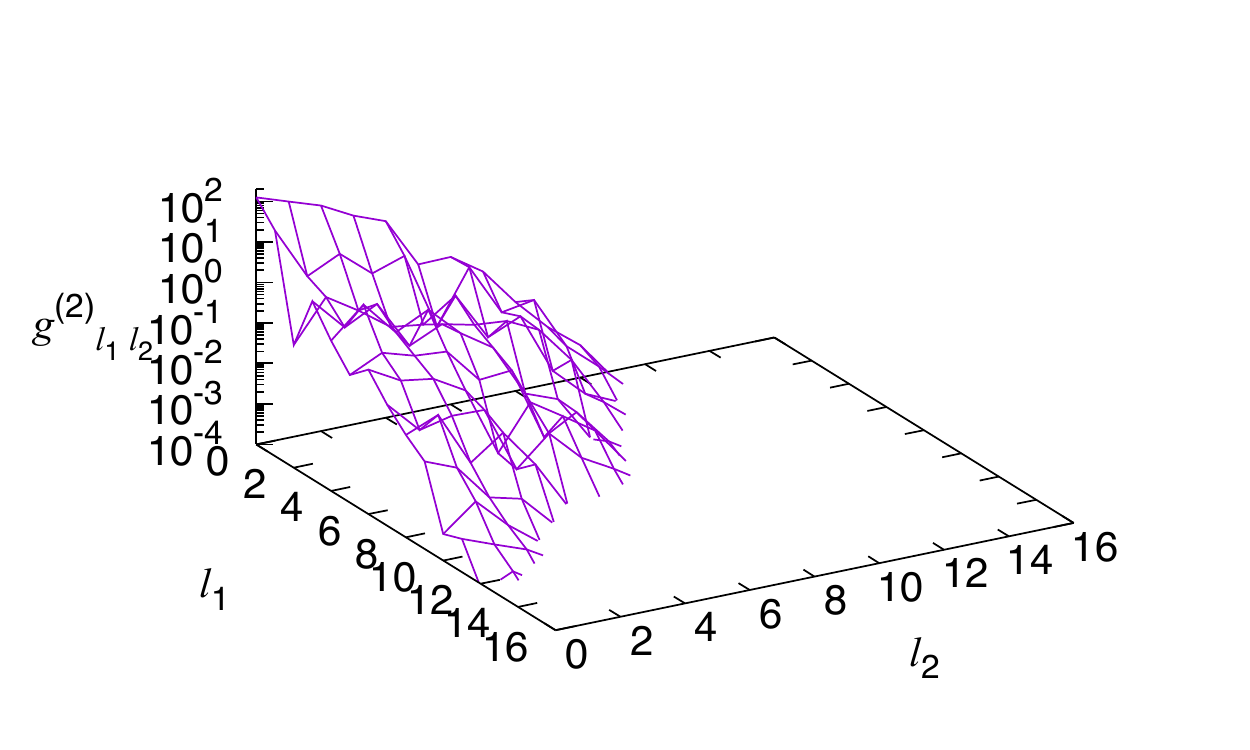}
	\includegraphics[width=0.4\textwidth,clip]{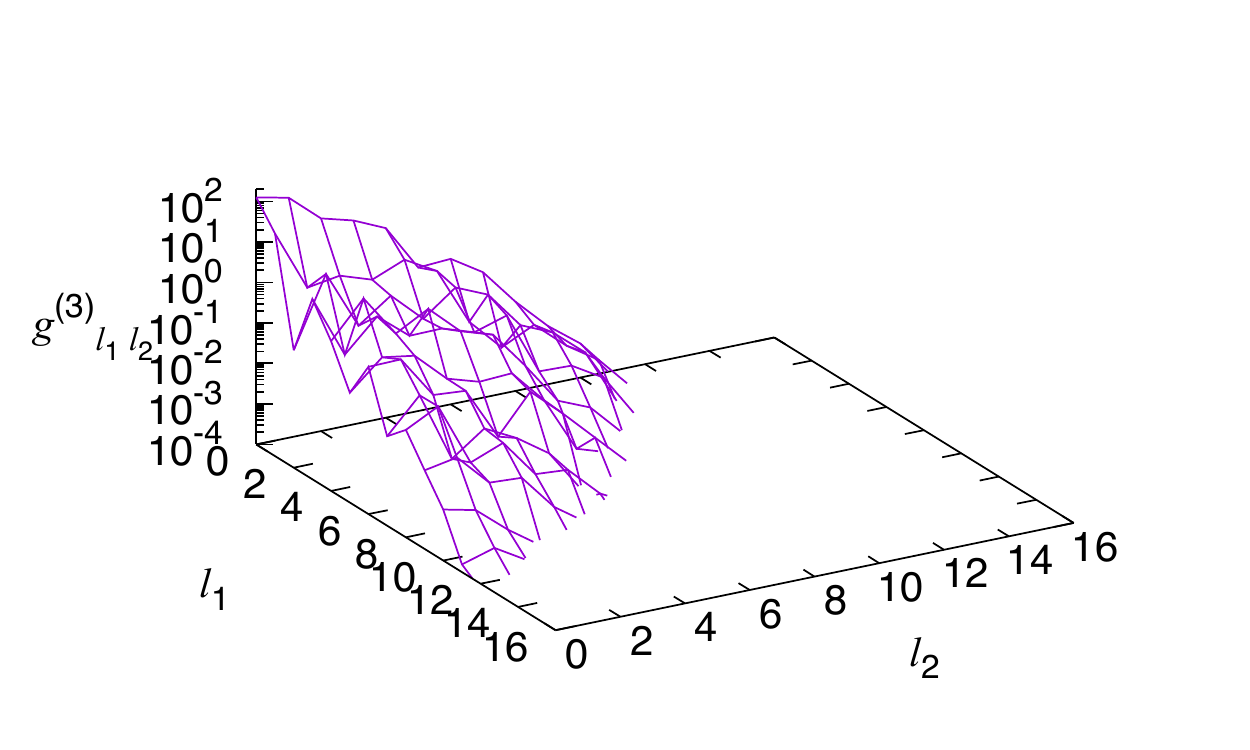}
	
	\caption{
		(Color online) Expansion coefficients obtained from fitting the QMC data shown in Fig.~\ref{fig:qmc}.
	}
	\label{fig:three-point-coeff-qmc}
\end{figure}
In this section, we demonstrate how to interpolate QMC data in Matsubara frequencies.
To this end, we solved the Hubbard model 
with semicircular non-interacting density of states of bandwidth 2
at half filling for $U=2$ and $\beta=20$.
The Hubbard model was solved with the dynamical mean-field approximation.
We solved quantum impurity problems by means of continuous-time hybridization expansion QMC method.\cite{Werner:2006ko}
For the converged solution,
we measured the three-point Green's function for the particle-hole channel:
\begin{equation}\begin{split}
&G^{\mathrm{ph}}_{\uparrow\uparrow}(i\nu_n, i\omega_{n^\prime}) \\
&\quad=\int_0^\beta d\tau_{12}\ d\tau_{23}\ e^{i\nu_n \tau_{12} + i\omega_{n^\prime} \tau_{23}}\ G^{\mathrm{ph}}_{\uparrow\uparrow}(\tau_1, \tau_2, \tau_3),
\end{split}\label{eq:Gph}
\end{equation}
where
\begin{align}
	G^{\mathrm{ph}}_{\uparrow\uparrow}(\tau_1, \tau_2, \tau_3) &= \braket{T_\tau c_\uparrow(\tau_1) c^\dagger_\uparrow(\tau_2) c_\uparrow(\tau_3) c_\uparrow^\dagger(\tau_3)}.\label{eq:Gph2}
\end{align}
Here, $\nu_n, \omega_{n^\prime}$ are fermionic and bosonic Matsubara frequencies, respectively.
We measured $G^{\mathrm{ph}}_{\uparrow\uparrow}(i\nu_n, i\omega_{n^\prime})$ in the rectangular Matsubara-frequency domain of $-100 \le n \le 99$ and $-100 \le n^\prime \le 100$ using worm improved estimators~\cite{Gunacker:2016vw}.
Recasting Eq.~(\ref{eq:Gph}) into the form of Eq.~(\ref{eq:three-point-exp2}),
we obtain the expansion formula 
\begin{align}
	&G^{\mathrm{ph}}(i\nu_n, i\omega_{n^\prime})\nonumber \\
	&=\int_0^\beta d\tau_1 d\tau_2 e^{i\nu_n \tau_{13} + i(\omega_{n^\prime}-\nu_n) \tau_{23}} G^{\mathrm{ph}}(\tau_1, \tau_2, \tau_3)\\
	&=  \sum_{l_1,l_2=0}^\infty \Big\{G^{(1)}_{l_1l_2}\UF_{l_1}(i\nu_n)\UF_{l_2}(i \omega_{n^\prime} - i \nu_n) \nonumber\\
	& \quad+  G_{l_1 l_2}^{(2)} \enhUB_{l_1}(i\omega_{n^\prime})\UF_{l_2}(i \omega_{n^\prime} - i \nu_n)+
	G_{l_1 l_2}^{(3)} \UF_{l_1}(i\nu_n)\enhUB_{l_2}(i\omega_{n^\prime})
	\Big\}.\label{eq:Gph-exp}
\end{align}
We fit the QMC data by this formula using the cost function in the form of 
Eq.~(\ref{eq:Ridge-Gthree}).
In this case, $\boldsymbol{G}^\mathrm{3pt} $ is a vector of length $2N_\mathrm{smp}$ consisting of the real and imaginary parts of the QMC data in the Matsubara frequency domain.
The matrix $\boldsymbol{A}$ is a matrix of size $(2N_\mathrm{smp}\times3N_\mathrm{IR}^2)$ that stores the real and imaginary parts of the coefficients in Eq.~(\ref{eq:Gph-exp}).
We use the same basis function as for the analysis in the previous section ($\beta \wmax = 40$ and $N_\mathrm{IR}=16$).
We set the regularization parameter to $\lambda = 10^{-8}$~\footnote{The regularization avoids overfitting to the QMC data with statistical errors.}.

The result is shown in Fig.~\ref{fig:qmc}.
One can see that the QMC data fit well within statistical errors by the present expression.
In particular, long-tail structures at the diagonal line $n=n^\prime$ in Fig.~\ref{fig:qmc}(a) are correctly described.
At the same time, the $\delta$-function-like feature at low frequency is reproduced. 

Figure~\ref{fig:three-point-coeff-qmc} shows the parameters obtained from fitting the QMC data.
It is clearly seen that the parameters decay exponentially at large $l_1$ and $l_2$ despite the presence of statistical errors.
As a consequence, the QMC data are interpolated smoothly without overfitting to statistical errors. 
These results indicate that the projection of QMC data to the present compact representation acts as a \textit{physically designed} noise filter.

\section{Summary and discussion}\label{sec:summary}
In summary, we derived an overcomplete non-orthogonal representation for two-particle Green's functions.
The basis functions for four (three) point response functions consist of 16 (4) combinations of products of 3 (2) single-particle IR basis functions.
The IR of the two-particle Green's functions successfully describes all discontinuities of two-particle Green's functions in imaginary time. These discontinuities are responsible for the high-frequency asymptotic behavior.
The delta-function-like low-frequency structures, which may arise from disconnected diagrams, are also described correctly in the same representation.
We showed rigorously that our expansion coefficients decay exponentially.
Our formalism allows the accurate and compact description of complex structures of the two-particle Green's functions.

Furthermore, an upper bound for the number of basis functions is known beforehand and can be derived from temperature and the energy scale of excitations.
Note that the number of basis functions for the single-particle Green's function grows only logarithmically with inverse temperature $\beta$.\cite{Chikano-unpublished}
This implies that the memory size required for storing the four-point Green's function increases more slowly than any power of $\beta$, which is a big advantage over the existing technologies.
For instance, one may be able to construct an overcomplete reprensentation of the two-particle Green's function in terms of Legendre polynomials (e.g., by expanding each term in Eq.~(\ref{eq:three-point-spectral-tau})).
This will also yield exponentially decaying expansion coefficients.
However, the result in Ref.~\onlinecite{Chikano-unpublished} indicates that the memory size required for storing the two-particle Green's function increases more rapidly as powers of $\beta$ for the Legendre representation.

In Sec.~\ref{sec:fit}, we demonstrated the accuracy of the present representations for the Hubbard atom.
In Sec.~\ref{sec:qmc}, we further tested the formulas using dynamical mean-field calculations of the single-site Hubbard model.
We also showed that the QMC data of the three-point Green's functions can be fitted with our expansion formula.
These results indicate that the expansion formulas can be used for interpolating QMC data. 

The present compact representation will open up new and interesting research applications. 
For example, efficient QMC measurement based on the compact representation will be useful especially for the four-point Green's function.
It will also enable implementation of the diagrammatic Monte Carlo methods, which can now be based on the multi-particle building blocks, and make the solution of the parquet equations more efficient.
\begin{acknowledgments}
	We are grateful to Lewin Boehnke, Giuseppe Carleo, Li Huang, Hugo Strand, Shintaro Hoshino, Philipp Werner, Masahiro Hasegawa and Takeo Kato for fruitful discussions.
	HS was supported by JSPS KAKENHI Grant No. 16H01064 (J-Physics), 16K17735.
	JO was supported by JSPS KAKENHI Grant No. 26800172, 16H01059 (J-Physics).
	MO was supported by MEXT KAKENHI Grant No. 25120008, JST CREST and JSPS KAKENHI No. 16H04382.
	KY was supported by Building of Consortia for the Development of Human Resources in Science and Technology, MEXT, Japan.
	KH was supported by NSF DMR-1405303.
	MO was supported by MEXT KAKENHI Grant No. 25120008, JSPS KAKENHI Grant No. 16H04382 and ImPACT Program of Council for Science, Technology and Innovation (Cabinet Office, Government of Japan). 
        MW was supported by the Simons collaboration on the many-electron problem, EG by NSF DMR-1606348.
    We used w2dynamics~\cite{w2dynamics} for QMC measurement of the three-point Green's function.
\end{acknowledgments}

\appendix

\section{Matsubara representation of the kernel}\label{appendix:kernel}
The transformation of $G^{\alpha}(\tau)$ to the Matsubara frequency domain is defined as
\begin{eqnarray}
	G^{\alpha}(i\omega_n) &=& \int_0^\beta \mathrm{d}\tau e^{\mathrm{i}\omega_n \tau} G^{\alpha}(\tau) \equiv \mathcal{F}(G^{\alpha}(\tau) ),
\end{eqnarray}
where $\omega_n = (2n+1)\pi/\beta$ in the fermionic case and $\omega_n = 2n\pi/\beta$ in the bosonic case, respectively.
$\mathcal{F}$ is the fourier transformation operator.

Equation~(\ref{eq:fwd}) can be reformulated as
\begin{align}
	G^{\alpha}(i\omega_n) &= \int_{-\infty}^{\infty} d\omega K^{\alpha}(i\omega_n, \omega) \rho^{\alpha}(\omega),\label{eq:fwd-iwn}
\end{align}
where 
\begin{eqnarray}
	\KF(i\omega_n, \omega) &\equiv& -\mathcal{F}(\KF(\tau, \omega)) = \frac{1}{i\omega_n - \omega},\\
	\KB(i\omega_n, \omega) &\equiv& -\mathcal{F}(\KB(\tau, \omega)) =\frac{\omega}{i\omega_n - \omega}.\label{eq:KB}
\end{eqnarray}

\begin{widetext}
\section{Decomposition of three-point Green's function}\label{appendix:three-point}
In this Appendix, we derive the spectral representation of the three-point Green's function, Eq.~(\ref{eq:three-point-spectral-tau}).
We start our discussion with the Fourier transform of $\Gthree(\tau_1, \tau_2, \tau_3)$ in Eq.~(\ref{eq:3point}). It reads
\begin{align}
	\Gthree(i\omega_1, i\omega_2) &= \int_0^\beta d\tau_1 \int_0^\beta d\tau_2 e^{i\omega_1 \tau_1 + i\omega_2 \tau_2} \Gthree(\tau_1, \tau_2, 0),
\label{eq:three-point-Fourier}
\end{align}
where $\omega_1$ and $\omega_2$ are fermionic Matsubara frequencies.
Here we have used the time-translation invariance $\Gthree(\tau_1, \tau_2, \tau_3) = \Gthree(\tau_{13}, \tau_{23}, 0)$ with $\tau_{ij} \equiv \tau_i - \tau_j$.
We introduce the sum over eigenstates with a special care on the contributions of zero bosonic excitation energies $E_{ik}=0$ to zero bosonic frequency $i\omega_1 + i\omega_2 = 0$.
One obtains
\begin{align}
	& \Gthree(i\omega_1, i\omega_2) \nonumber\\
	&= Z^{-1}\sum_{ijk} \Big\{e^{-\beta E_i} \frac{A_{ij}B_{jk}C_{ki}}{i\omega_2 + E_{jk}}
	\left[
	  \frac{e^{\beta E_{ik}}-1}{i\omega_1 + i\omega_2 + E_{ik}} (1-\delta_{E_{ik},0})
	  + \beta \delta_{E_{ik},0}\delta_{\omega_1+\omega_2,0}
	  + \frac{e^{\beta E_{ij}}+1}{i\omega_1 +  E_{ij}}
	\right]\nonumber\\
	&\hspace{4.5em} - e^{-\beta E_i} \frac{B_{ij}A_{jk}C_{ki}}{i\omega_1 + E_{jk}}
	\left[
	 \frac{e^{\beta E_{ik}}-1}{i\omega_1 + i\omega_2 + E_{ik}} (1-\delta_{E_{ik},0})
	 + \beta \delta_{E_{ik},0}\delta_{\omega_1+\omega_2,0}
	 + \frac{e^{\beta E_{ij}}+1}{i\omega_2 +  E_{ij}}
	\right]\Big\}.\label{eq:three-point-lehmann}
\end{align}
Here, we have defined $E_{\alpha\beta} \equiv E_\alpha - E_\beta$,
where $\alpha$ and $\beta$ denote eigenstates.
We have also defined 
$A_{\alpha\beta} \equiv \braket{\alpha|A|\beta}$,
$B_{\alpha\beta} \equiv \braket{\alpha|B|\beta}$,
$C_{\alpha\beta} \equiv \braket{\alpha|C|\beta}$.
We now split $\Gthree$ into a singular part $\Gthreesingular$ and a normal part $\Gthreenorm$ as
\begin{align}
	& \Gthree(i\omega_1, i\omega_2) = \Gthreenorm(i\omega_1, i\omega_2) + \Gthreesingular(i\omega_1, i\omega_2),\\
	& \Gthreenorm(i\omega_1, i\omega_2) \equiv Z^{-1}\sum_{ijk} (e^{-\beta E_i}+e^{-\beta E_j})
	\Big\{ \frac{A_{ij}B_{jk}C_{ki}}{(i\omega_1 +  E_{ij})(i\omega_2 + E_{jk})}
	  -  \frac{B_{ij}A_{jk}C_{ki}}{(i\omega_1 + E_{jk})(i\omega_2 +  E_{ij})}
	\Big\}\nonumber\\
	&\hspace{2em}
	    -Z^{-1}\sum_{ijk, E_i\neq E_k}(e^{-\beta E_k}-e^{-\beta E_i})
	    \left\{
	        \frac{A_{ij}B_{jk}C_{ki}}{(i\omega_2 + E_{jk})(i\omega_1 + i\omega_2 + E_{ik})}
	        -
	        \frac{B_{ij}A_{jk}C_{ki}}{(i\omega_1 + E_{jk})(i\omega_1 + i\omega_2 + E_{ik})}
	    \right\}
	,\label{eq:three-point-lehmann-norm}\\
	& \Gthreesingular(i\omega_1, i\omega_2) \equiv \beta\delta_{\omega_1+\omega_2,0} Z^{-1}\sum_{ijk, E_i=E_k} e^{-\beta E_i} \Big\{-\frac{A_{ij}B_{jk}C_{ki}}{i\omega_1 - E_{jk}}
    -\frac{B_{ij}A_{jk}C_{ki}}{i\omega_1 + E_{jk}}
	\Big\}.
\end{align}

The normal part $\Gthreenorm$ can readily be recast into
\begin{align}
	&\Gthreenorm(i\omega_1, i\omega_2)\nonumber\\
	&= \int_{-\infty}^\infty d\epsilon_1 d \epsilon_2
	\Big[ 
	\frac{\rho^{(1)}(\epsilon_1, \epsilon_2)}{(i\omega_1 - \epsilon_1)(i\omega_2 - \epsilon_2)}+
	\frac{\epsilon_1 \rho^{(2)}(\epsilon_1, \epsilon_2)}{(i\omega_1+i\omega_2 - \epsilon_1)(i\omega_2 - \epsilon_2)}+
	\frac{\epsilon_1 \rho^{(3)}(\epsilon_1, \epsilon_2)}{(i\omega_1+i\omega_2 - \epsilon_1)(i\omega_1 - \epsilon_2)}
	\Big],\nonumber\\
    & = \int_{-\infty}^\infty d\epsilon_1 d \epsilon_2
    \Big[
    \KF(i\omega_1, \epsilon_1)\KF(i\omega_2, \epsilon_2) \rho^{(1)}(\epsilon_1, \epsilon_2)+
    \KB(i\omega_1+i\omega_2, \epsilon_1)\KF(i\omega_2, \epsilon_2) \rho^{(2)}(\epsilon_1, \epsilon_2) \nonumber \\
    & \hspace{2em}+ \KB(i\omega_1+i\omega_2, \epsilon_1)\KF(i\omega_1, \epsilon_2) \rho^{(3)}(\epsilon_1, \epsilon_2)
    \Big].
	~\label{eq:three-point-spectral}
\end{align}
The Matsubara frequency representation of the kernel $K^\alpha(i\omega_n, \omega)$ is defined in Appendix~\ref{appendix:kernel}.
We have introduced three distinct spectral functions
$\rho^{(1)}(\epsilon_1, \epsilon_2)$,
$\rho^{(2)}(\epsilon_1, \epsilon_2)$,
$\rho^{(3)}(\epsilon_1, \epsilon_2)$:
\begin{align}
	\rho^{(1)}(\epsilon_1, \epsilon_2) &\equiv Z^{-1} \sum_{ijk} (e^{-\beta E_i}+e^{-\beta E_j})\Big[
	A_{ij} B_{jk} C_{ki} \delta(\epsilon_1+E_{ij})\delta(\epsilon_2+E_{jk})-
	B_{ij} A_{jk} C_{ki}\delta(\epsilon_1+E_{jk})\delta(\epsilon_2+E_{ij})
	\Big],\\
	\rho^{(2)}(\epsilon_1, \epsilon_2) &\equiv  Z^{-1}\epsilon_1^{-1}\sum_{ijk, E_i\neq E_k} A_{ij} B_{jk} C_{ki} (e^{-\beta E_k}-e^{-\beta E_i}) \delta(\epsilon_1+E_{ik})\delta(\epsilon_2+E_{jk}),\label{eq:rho2}\\
	\rho^{(3)}(\epsilon_1, \epsilon_2) &\equiv  -Z^{-1}\epsilon_1^{-1}\sum_{ijk, E_i\neq E_k} B_{ij}A_{jk}C_{ki} (e^{-\beta E_k}-e^{-\beta E_i}) \delta(\epsilon_1+E_{ik})\delta(\epsilon_2+E_{jk}).\label{eq:rho3}
\end{align}
The inverse Fourier transform of Eq.~(\ref{eq:three-point-spectral}) reads
\begin{align}
	\Gthreenorm(\tau_1, \tau_2, 0) 
	& = \int d\epsilon_1 d \epsilon_2 \KF(\tau_1,\epsilon_1) \KF(\tau_2,\epsilon_2)\rho^{(1)}(\epsilon_1,\epsilon_2)+  \KB(\tau_1,\epsilon_1) \KF(\tau_{21},\epsilon_2)\rho^{(2}(\epsilon_1,\epsilon_2)\nonumber\\
	&\hspace{2em}+  \KF(\tau_{12},\epsilon_1) \KB(\tau_2,\epsilon_2)\rho^{(3)}(\epsilon_1,\epsilon_2),\label{eq:Gthree-tau1-tau2-Kernel}
\end{align}
Here, we have used the following relation between different notations of imaginary time:
\begin{align}
	\int_0^\beta d\tau_1 d\tau_2 \Gthree(\tau_1, \tau_2,0)e^{i\omega_1 \tau_1 + i \omega_2\tau_2}
	&= \int_0^\beta d\tau_1 d\tau_{21}  \Gthree(\tau_1, \tau_2,0) e^{(i\omega_1+i\omega_2) \tau_1 + i \omega_2 \tau_{21}}
	\nonumber\\
	&=  \int_0^\beta d\tau_1 d\tau_{12}  \Gthree(\tau_1, \tau_2,0)e^{(i\omega_1+i\omega_2) \tau_2 + i \omega_1 \tau_{12}}.\label{eq:Gthree-FT}
\end{align}
Replacing $\tau_1$ with $\tau_{13}$ and $\tau_2$ with $\tau_{23}$ in Eq.~(\ref{eq:Gthree-tau1-tau2-Kernel}), we obtain the first three terms in Eq.~(\ref{eq:three-point-spectral-tau}).

The remaining task is to analyze the singular term:
\begin{align}
	\Gthreesingular(i\omega_1, i\omega_2)
	&= \beta \delta_{\omega_1+\omega_2,0} \int d\epsilon \frac{\rho_\mathrm{singular}(\epsilon)}{i\omega_1-\epsilon},
	\nonumber\\
	&= \beta \delta_{\omega_1+\omega_2,0} \int d\epsilon \KF(i\omega_1,\epsilon) \rho_\mathrm{singular}(\epsilon),
	\label{eq:three-point-singular}
\end{align}
where
\begin{align}
    \rho_\mathrm{singular}(\epsilon) 
    = -Z^{-1}\sum_{ijk, E_i=E_k} e^{-\beta E_i}
     \big[ A_{ij}B_{jk}C_{ki}\delta(\epsilon-E_{ji}) + B_{ij}A_{jk}C_{ki}\delta(\epsilon+E_{ji}) \big].~\label{eq:three-point-singular-rho}
\end{align}
Using Eq.~(\ref{eq:Gthree-FT}), one obtains the inverse Fourier transform of Eq.~(\ref{eq:three-point-singular}) as
\begin{align}
\Gthreesingular(\tau_{12})
&= \int d\epsilon
K^\mathrm{F}(\tau_{12}, \epsilon)\rho_\mathrm{singular}(\epsilon).~\label{eq:three-point-singular-tau2}
\end{align}
The last term in Eq.~(\ref{eq:three-point-spectral-tau}) is thus derived.

\section{Decomposition of four-point Green's function}\label{appendix:four-point}
In this Appendix, we derive the decomposition formula, Eq.~(\ref{eq:ir-four-point}), for the four-point Green's function $\Gfour(\tau_1, \tau_2, \tau_3, \tau_4)$.
We start from the spectral representation, which has been derived in Refs.~\onlinecite{Toschi07,Hafermann09b,Shvaika:2016gf}.
The expression presented in Ref.~\onlinecite{Hafermann09b} reads
\begin{align}
& \Gfour(i\omega_1, i\omega_2, i \omega_3, i \omega_4 ) \nonumber \\
& = \int_0^\beta d\tau_1 d\tau_2 d\tau_3 d\tau_4 \braket{T_\tau A(\tau_1) B(\tau_2) C(\tau_3) D(\tau_4)} e^{i\omega_1 \tau_1 + i\omega_2 \tau_2 + i\omega_3 \tau_3 + i \omega_4 \tau_4} \nonumber\\
& = Z^{-1} \beta \delta_{\omega_1 + \omega_2 + \omega_3 + \omega_4,0}
\sum_{\Pi} \mathrm{sgn(\Pi)}
\sum_{ijkl} 
\langle i | O_{\Pi_1} | j\rangle
\langle j | O_{\Pi_2} | k\rangle
\langle k | O_{\Pi_3} | l\rangle
\langle l | O_{4} | i\rangle
\phi(E_i, E_j, E_k, E_l, \omega_{\Pi_1}, \omega_{\Pi_2}, \omega_{\Pi_3}),\label{eq:four-point-spectral-tau4}
\end{align}
where
\begin{align} 
	\phi(E_i, E_j, E_k, E_l, \omega_1, \omega_2, \omega_3) &\equiv \int_0^\beta d\tau_1 \int_0^{\tau_1} d\tau_2 \int_0^{\tau_2} d \tau_3
	e^{-\beta E_i}e^{E_{ij}\tau_1} e^{E_{jk}\tau_2}e^{E_{kl}\tau_3}e^{i(\omega_1\tau_1 + \omega_2\tau_2 + \omega_3\tau_3)}.\label{eq:phi}
\end{align}
To deal with the four operators on equal footing,
we keep four frequencies in the equations explicitly.
As in Ref.~\onlinecite{Hafermann09b}, we have defined $O_1 = A$, $O_2 = B$, $O_3=C$, and $O_4=D$.
We take summation over all permutations $\Pi$ of the indices ${123}$.
The integrals in Eq.~(\ref{eq:phi}) can be evaluated explicitly by taking special care of contributions arising from zero excitation energies. 
We further split the result into three terms:
\begin{align}
&\phi(E_i, E_j, E_k, E_l, \omega_1, \omega_2, \omega_3)
= \sum_{n=1}^{3} \phi^{(n)}(E_i, E_j, E_k, E_l, \omega_1, \omega_2, \omega_3),
\end{align}
where
\begin{align}
&\phi^{(1)}(E_i, E_j, E_k, E_l, \omega_1, \omega_2, \omega_3)\equiv
\frac{e^{-\beta E_i} + e^{-\beta E_j}}{(i\omega_3 + E_{kl})(i\omega_2 + E_{jk})(i\omega_1 + E_{ij})},\\
&\phi^{(2)}(E_i, E_j, E_k, E_l, \omega_1, \omega_2, \omega_3)\equiv
\frac{1}{(i\omega_3 + E_{kl})} \Bigg[
\frac{1-\delta_{\omega_2+\omega_3,0}\delta_{E_j,E_l}}{i\omega_2 + i\omega_3 + E_{jl}}
\left\{
\frac{e^{-\beta E_i} + e^{-\beta E_j} }{i\omega_1 + E_{ij}} - \frac{e^{-\beta E_i} + e^{-\beta E_l}}{i\omega_1 + i\omega_2 + i\omega_3 + E_{il}} \right\} + \nonumber\\
& \quad \delta_{\omega_2+\omega_3,0}\delta_{E_j,E_l}
\left(
    \frac{e^{-\beta E_i}+e^{-\beta E_j}}{(i\omega_1 + E_{ij})^2}
    -\beta\frac{e^{-\beta E_j}}{i\omega_1 + E_{ij}}
\right)
\Bigg],
\\
&\phi^{(3)}(E_i, E_j, E_k, E_l, \omega_1, \omega_2, \omega_3) \equiv
\frac{1}{(i\omega_3 + E_{kl})} \Bigg[
- \frac{1}{i\omega_2 + E_{jk}}
    \left(
        \frac{e^{-\beta E_i} + e^{-\beta E_j}}{i\omega_1 + E_{ij}} -
         (1-\delta_{\omega_1+\omega_2,0}\delta_{E_i,E_k})\frac{e^{-\beta E_i} -  e^{-\beta E_k}}{i\omega_1 + i\omega_2 + E_{ik}}
    \right)\nonumber\\
    &\quad + \beta e^{-\beta E_i} \delta_{\omega_1+\omega_2,0}\delta_{E_i,E_k}
\Bigg].
\end{align}

\begin{table}
	\centering
	\begin{tabular}{c|c}
		\hline
		($i\omega$, $i\omega$', $i\omega$'') & ($\tau$, $\tau'$, $\tau''$)\\
		\hline
		\hline
		($i\omega_3$, $i\omega_2+i\omega_3$, $-i\omega_1$) & ($\tau_{32}$, $\tau_{24}$, $\tau_{41}$) \\
		($i\omega_3$, $i\omega_2+i\omega_3$, $-i\omega_4=i\omega_1+i\omega_2+i\omega_3$) & ($\tau_{32}$, $\tau_{21}$, $\tau_{14}$) \\
	   ($i\omega_3$, $i\omega_2$, $i\omega_1$) & ($\tau_{34}$, $\tau_{24}$, $\tau_{14}$) \\
	   ($-i\omega_3$, $i\omega_2$, $i\omega_1+i\omega_2$) & ($\tau_{43}$, $\tau_{21}$, $\tau_{14}$) \\
		\hline
	\end{tabular}
	\caption{Mapping between Matsubara frequencies and imaginary times used for decomposing $\Gfour$.}
	\label{table:G4pt}
\end{table}

Now we perform the inverse Fourier transformation using notations of frequencies in Table~\ref{table:G4pt}.
The result reads
\begin{align}
	&\phi^{(1)}(E_i, E_j, E_k, E_l, \tau_1, \tau_2, \tau_3, \tau_4) \equiv f(E_i,E_j)\KF(\tau_{14},-E_{ij})\KF(\tau_{24},-E_{jk})\KF(\tau_{34},-E_{kl}),\label{eq:phi1-tau}\\
	&\phi^{(2)}(E_i, E_j, E_k, E_l, \tau_1, \tau_2, \tau_3, \tau_4)=\nonumber\\
	&\hspace{1em}
	\begin{cases}
		-\frac{\KF(\tau_{32}, -E_{kl})}{E_{jl}}\Big(
		f(E_i,E_j) \KB(\tau_{24}, -E_{jl}) \KF(\tau_{41}, E_{ij}) 
		-f(E_i,E_l)\KB(\tau_{21}, -E_{jl}) \KF(\tau_{41}, E_{il}) 
		\Big) & (E_{jl} \neq 0),\\
		\beta^{-1}f(E_i,E_j)\KF(\tau_{32}, -E_{kl}) \KF(\tau_{41}, E_{ij}) \Big\{
		-T(\tau_{41})+ T(\tau_{42}) +  \beta + T(\tau_{21}) \Big\}
		& (E_{jl}= 0)
	\end{cases},\label{eq:phi2-tau}\\
	&\phi^{(3)}(E_i, E_j, E_k, E_l, \tau_1, \tau_2, \tau_3, \tau_4)=\nonumber\\
	& \hspace{1em}
	\begin{cases}
		-\frac{e^{-\beta E_i} - e^{-\beta E_k}}{E_{ik}}  \KF(\tau_{21},-E_{jk})\KB(\tau_{14}, -E_{ik}) \KF(\tau_{43}, E_{kl})
		& (E_{ik}\neq 0)\\
		e^{-\beta E_i}\KF(\tau_{21},-E_{jk}) \KF(\tau_{43}, E_{kl}) & (E_{ik}= 0)
	\end{cases},\label{eq:phi3-tau}
\end{align}
where $T(\tau) \equiv \tau$ for $0 < \tau < \beta$ with the periodicity of $T(\tau+\beta) = T(\tau)$, and $f(E,E^\prime) \equiv e^{-\beta E} + e^{-\beta E^\prime}$.
The factors $1/E_{jl}$ and $1/E_{ik}$ in Eqs.~(\ref{eq:phi2-tau}) and (\ref{eq:phi3-tau}) originate from the extra $\omega$ in the numerator of $\KB(i\omega_n, \omega)$ [see Eq.~(\ref{eq:KB})].
To derive Eqs.~(\ref{eq:phi1-tau})--(\ref{eq:phi3-tau}),
we used
\begin{align}
	\mathcal{F}[T(\tau)] &= \beta \frac{1-\delta_{n,0}}{i\nu_n} + \frac{1}{2}\beta^2,\\
	\mathcal{F}[\KF(\tau,\epsilon)(T(\tau)-\beta n^\mathrm{F}(\epsilon))] &= \frac{1}{(i\omega_n-\epsilon)^2},\\
	n^\mathrm{F}(\epsilon) &\equiv \frac{1}{1+e^{\beta \epsilon}},
\end{align}
where $\nu_n$ is a bosonic Matsubara frequency and $\omega_n$ is a fermionic Matsubara frequency.
The term $\beta + T(\tau_{21})-T(\tau_{24})-T(\tau_{41})$ in Eqs.~(\ref{eq:phi2-tau}) evaluates to a constant within each domain 
where time ordering is explicit (i.e. each tetrahedron shown in Fig.~\ref{fig:equal-planes}).
This term acts like a step function, playing an important role in describing  the discontinuities of the singular contributions.

Now we assign each term in Eqs.~(\ref{eq:phi1-tau})--(\ref{eq:phi3-tau}) to one of the 16 representations in Table~\ref{table:summary}.
The results are summarized in Fig.~\ref{fig:G4pt-diagram-appendix}.
We used the relation $K^\alpha(\tau,\omega) = \mp K^\alpha(-\tau,-\omega)$ ($\alpha$=F,~B) and defined the identity function $I(\tau) = 1$ with te bosonic statistics.
The derivation is the same as $\Gthree$ except for the first term in Eq.~(\ref{eq:phi2-tau}) ($E_{jl} = 0$), which depends on $\tau_{41}$ as $T(\tau_{41}) \KF(\tau_{41}, E_{ij})$.
This term may appear in the spectral representation as
\begin{align}
& \int_{-\infty}^\infty d\epsilon_1 d\epsilon_2 \KF(\tau_{32}, \epsilon_1) T(\tau_{41})\KF(\tau_{41}, \epsilon_2) \rho(\epsilon_1, \epsilon_2)\nonumber \\
\propto &\sum_{l_1,l_2} G_{l_1,l_2}\UF_{l_1}(\tau_{32})T(\tau_{41})\UF_{l_2}(\tau_{41}),\label{eq:tau-corr}
\end{align}
where $G_{l_1,l_2}$ decay as $\propto s_{l_1} s_{l_2}$ when $\beta\wmax$ is large enough.
We numerically found that $T(\tau)\UF_{l}(\tau)$, which has fermionic statistics, is expanded as a linear combination of $\UF_{l-1}(\tau)$ and $\UF_{l+1}(\tau)$ very precisely.
This is because the $\tau$ dependence of $T(\tau)\UF_{l}(\tau)$ is dominated by $\UF_{l}(\tau)$ at large $l$.
Thus, Eq.~(\ref{eq:tau-corr}) can be recast into
\begin{align}
&\sum_{l_1,l_2} \tilde{G}_{l_1,l_2}\UF_{l_1}(\tau_{32})\UF_{l_2}(\tau_{41})\label{eq:tau-corr2}
\end{align}
with $\tilde{G}_{l_1,l_2}$ decaying as $\propto s_{l_1} s_{l_2}$ at large $l_1$ and $l_2$.
This allows to assign this term to \# 9.

\begin{figure}
	\centering
	\includegraphics[width=0.4\textwidth,clip]{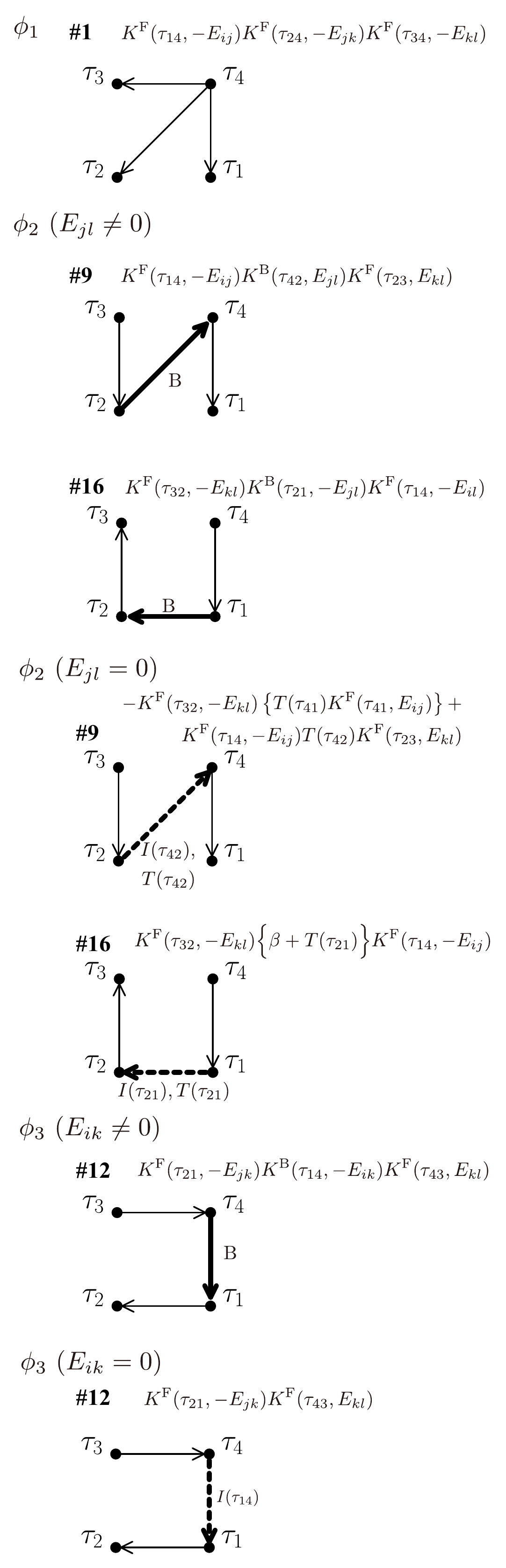}
	\caption{
		(Color online) Diagrams representing  Eqs.~(\ref{eq:phi1-tau})--(\ref{eq:phi3-tau}).
	}
	\label{fig:G4pt-diagram-appendix}
\end{figure}

Considering all permutations of $\tau_1,\tau_2,\tau_3$,
one can clarify all the terms of Eq.~(\ref{eq:four-point-spectral-tau4}) into 13 representations in Table.~\ref{table:summary}.
However, terms in the form of \#2, \#3, \#4 do not appear.
This is due to our choice of $\tau_4$ as the origin of imaginary time made in Eq.~(\ref{eq:four-point-spectral-tau4}).
After considering permutations including $\tau_4$,
we obtain the symmetrized 16 representations in Table~\ref{table:summary}.

\section{Expressions of $\Gthree$ for the Hubbard atom}\label{appendix:single-site}
We give the explicit expressions of $\Gthree$ for the Hubbard atom in Eq.~(\ref{eq:Hubbard-atom}) at half filling ($\mu=U/2$).
In the Matsubara-frequency domain, the two contributions to $\Gthree(i\omega_1, i\omega_2)$ is given by
\begin{align}
   \Gthreesingular (i\omega_1, i\omega_2)
   &= 
   \frac{\beta \delta_{\omega_1+\omega_2,0}}{2(1+e^{-\beta U/2})}
   \nonumber \\
   &\times
   \left(
     \frac{1}{i\omega_2 + U/2 }-
     \frac{e^{-\beta U/2}}{i\omega_2 - U/2 }
     \right),\\
   \Gthreenorm (i\omega_1, i\omega_2)
   &= \frac{1}{2}\Bigg[
    \frac{1}{(i\omega_1 - U/2)(i\omega_2 + U/2) }
   \nonumber \\
   &+
   \frac{1}{(i\omega_1 + U/2)(i\omega_2 - U/2) }  \Bigg].
\end{align}
These expressions are transformed into the imaginary-time domain to yield
\begin{align}
  &\Gthreesingular (\tau_{12})
  =
  -\frac{1}{2(1+e^{-\beta U/2})}
  \nonumber\\
  &\hspace{2em}\times
  \left[ K^\mathrm{F}(\tau_{21}, -U/2) - e^{-\beta U/2}K^\mathrm{F}(\tau_{21}, U/2) \right],\\
  &\Gthreenorm (\tau_1, \tau_2,0)
  =
  \frac{1}{2} \Big[ K^\mathrm{F}(\tau_1, U/2)K^\mathrm{F}(\tau_2, -U/2)
  \nonumber\\
  &\hspace{8em}+ K^\mathrm{F}(\tau_1, -U/2)K^\mathrm{F}(\tau_2, U/2) \Big]. 
\end{align}
Figure~\ref{fig:Gthree-decomposition} shows the contributions of $\Gthreenorm$ and $\Gthreesingular$ to the data in Fig.~\ref{fig:sampling-point} separately.
One can see substantial contributions from the singular term. 
\begin{figure}
	\centering
	\includegraphics[width=0.45\textwidth,clip]{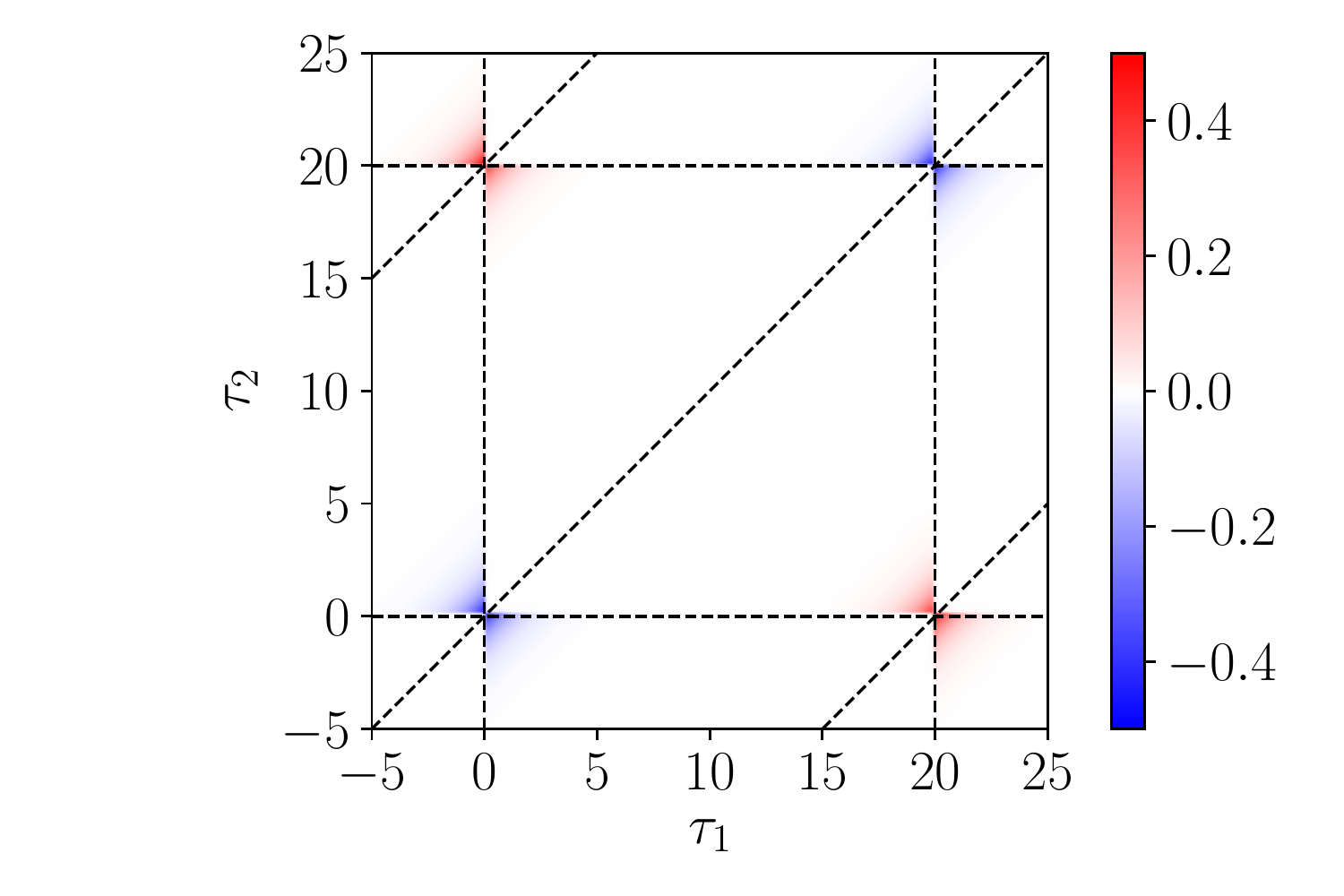}\\
	\includegraphics[width=0.45\textwidth,clip]{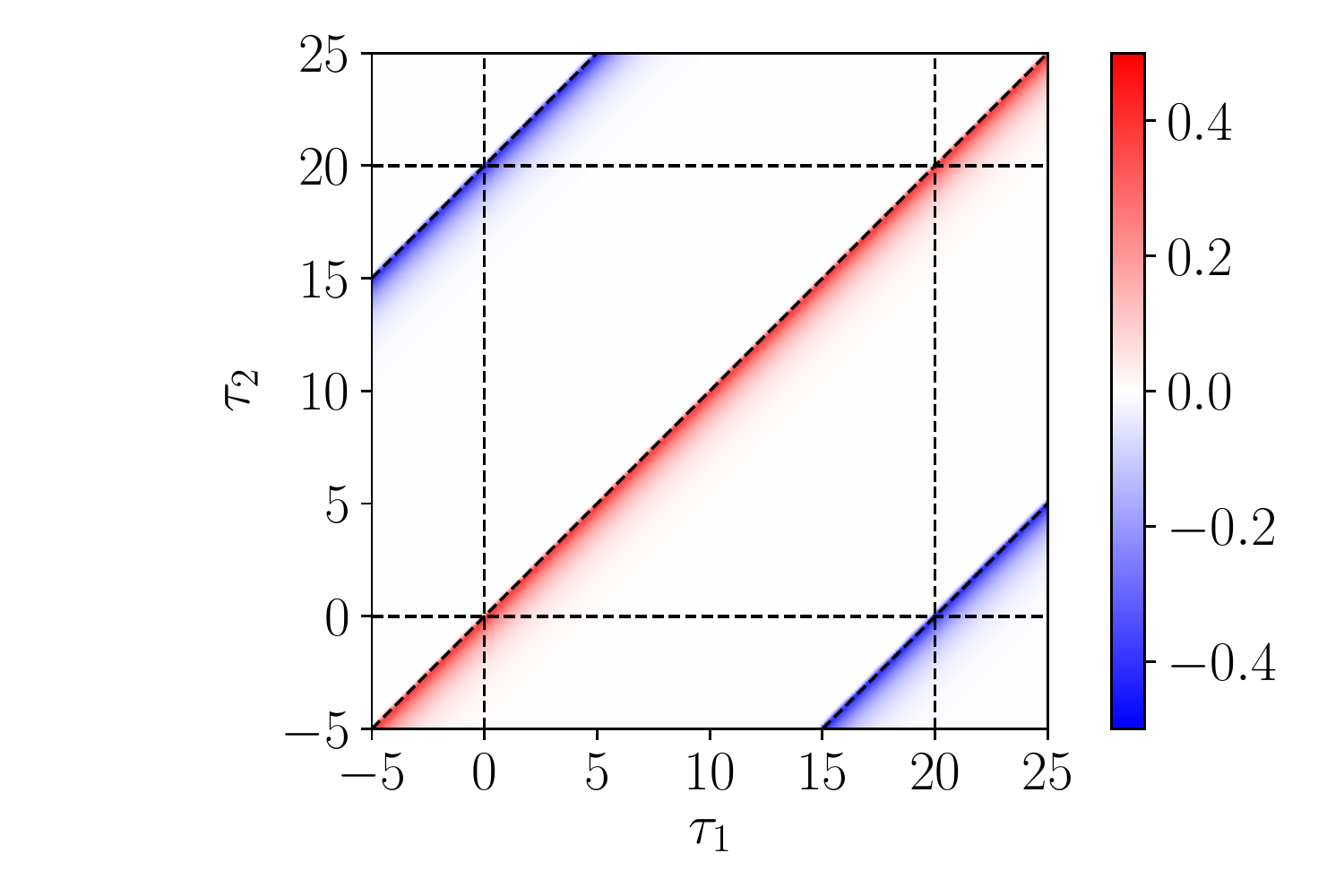}
	\caption{
		(Color online) Contributions of $\Gthreenorm$ (upper panel) and $\Gthreesingular$ (lower panel) to the data in Fig.~\ref{fig:sampling-point}.
	}
	\label{fig:Gthree-decomposition}
\end{figure}

\section{Ridge regression}\label{appendix:ridge}
The cost function of Ridge regression is given by
\begin{eqnarray}
||\boldsymbol{y} - \boldsymbol{A} \boldsymbol{x} ||^2+ \lambda ||\boldsymbol{x}||^2,\label{eq:Ridge}
\end{eqnarray}
where $\lambda~(>0)$ is a regularization parameter.
The solution of this minimization problem is given by the formula
\begin{eqnarray}
   \boldsymbol{x}^* &= (\boldsymbol{A}^T \boldsymbol{A}  + \lambda \boldsymbol{I} )^{-1} \boldsymbol{A} ^T \boldsymbol{y},
\end{eqnarray}
where $\boldsymbol{I}$ is an identity matrix.
The inverse always exists because $\boldsymbol{A}^T \boldsymbol{A}  + \lambda \boldsymbol{I}$ is positive definite: A positive $\lambda$ stabilizes the inversion.
In the present study, we used the implementation of Ridge regression in scikit-learn~\cite{scikit-learn}.

Equation (\ref{eq:Ridge-Gthree}) can be recast into the form of Eq.~(\ref{eq:Ridge}) by defining $\boldsymbol{x} \equiv ( G^{(r)}_{l_1 l_2}/S^{(r)}_{l_1 l_2} )$ and changing the definition of $\boldsymbol{A}$ accordingly.

\section{Non-uniform grids for $\Gfour$}\label{appendix:grid}
Let $\set{\tau^\alpha_1, \cdots, \tau^\alpha_{15}}$ be the 15 nodes of $U^\alpha_{15}(\tau)$ ($\alpha=\mathrm{F}$, B) in ascending order.
We define 16 sampling points as
$\boldsymbol{\tau}^\alpha \equiv \set{\tau^\alpha_1/2, (\tau^\alpha_1 +  \tau^\alpha_2)/2
	,\cdots, (\tau^\alpha_{15}+\beta)/2}$.
A non-uniform grid is then generated as a product of $\boldsymbol{\tau}^F$
with respect to $\tau_{14}$, $\tau_{24}$ and $\tau_{34}$ (i.e. in the representation \#1).
In a similar manner, two additional grids are generated for the representations \#2 and \#5, respectively.
We thus obtained $N_\mathrm{smp}=3\times 16^3$ sampling points which represent the complicated $\tau$'s dependence efficiently.

\end{widetext}

\bibliography{JO,ref,unpublished}
\end{document}